\RequirePackage[hyphens]{url} 
\documentclass[%
 reprint,
 superscriptaddress,
 showpacs,amsmath,amssymb,
 aps,
floatfix,
nofootinbib
]{revtex4-1}
\usepackage{url}
\usepackage[breaklinks]{hyperref}
\PassOptionsToPackage{hyphens}{url}\usepackage{hyperref}
\usepackage{mathpazo}
\usepackage{times}
\usepackage{graphicx}
\usepackage{array}
\usepackage{siunitx}
\usepackage{dcolumn}%
\usepackage{footnote}
\usepackage{makecell,multirow}
\usepackage{booktabs}
\usepackage[utf8]{inputenc}
\usepackage{tikz}
\usetikzlibrary{shapes,arrows}
\usepackage{multirow}
\usepackage{siunitx}
\usepackage{amsmath}
\usepackage{bm}
\usepackage[mathlines]{lineno}
\usepackage{braket}
\usepackage{color,soul}
\usepackage{subfigure}
\usetikzlibrary{shapes.geometric}
\usepackage{changepage}
\usepackage[bottom]{footmisc}
\usepackage{float}
\usepackage[american,siunitx]{circuitikz}
\usetikzlibrary{arrows,calc,positioning}
\usepackage{comment}

\newsavebox{\foobox}

\tikzset{ar/.style={-latex,shorten >=-1pt, shorten <=-1pt}}
\begin{document}
\preprint{APS/123-QED}

\title{Non-Virialized Axion Search Sensitive to Doppler Effects in the Milky Way Halo}
\author{C.Bartram}
\affiliation{University of Washington, Seattle, WA 98195, USA}
\affiliation{Stanford Linear Accelerator Center, Menlo Park, California, 94025, USA}
\author{T. Braine}
  \affiliation{University of Washington, Seattle, WA 98195, USA}
    \author{R. Cervantes}
    \affiliation{University of Washington, Seattle, WA 98195, USA}
  \affiliation{Fermi National Accelerator Laboratory, Batavia IL 60510, USA}
 \author{N. Crisosto}
 \email[Work was done prior to joining Amazon]{}{}
  \affiliation{University of Washington, Seattle, WA 98195, USA}
  \affiliation{currently at AWS Center for Quantum Computing, Pasadena, CA 91125}
  
  \author{N. Du}%
  \affiliation{University of Washington, Seattle, WA 98195, USA}
\affiliation{Lawrence Livermore National Laboratory, Livermore, CA 94550, USA}
\author{C. Goodman}
  \affiliation{University of Washington, Seattle, WA 98195, USA}
\author{M. Guzzetti}
\email[Correspondence to: ]{mguzz28@uw.edu and nicole.man@pnnl.gov}
\affiliation{University of Washington, Seattle, WA 98195, USA}
\author{C. Hanretty}
  \affiliation{University of Washington, Seattle, WA 98195, USA}
  \author{S. Lee}
  \affiliation{University of Washington, Seattle, WA 98195, USA}
  \author{G. Leum}
\affiliation{University of Washington, Seattle, WA 98195, USA}
\affiliation{currently at Microsoft Quantum, Redmond, WA 98052}
 \author{L. J Rosenberg}%
  \affiliation{University of Washington, Seattle, WA 98195, USA}
  \author{G. Rybka}%
  \affiliation{University of Washington, Seattle, WA 98195, USA}
  \author{J. Sinnis}
\affiliation{University of Washington, Seattle, WA 98195, USA}
\author{D. Zhang}
\affiliation{University of Washington, Seattle, WA 98195, USA}  

\author{M. H. Awida} 
  \affiliation{Fermi National Accelerator Laboratory, Batavia IL 60510, USA}
\author{D. Bowring}
  \affiliation{Fermi National Accelerator Laboratory, Batavia IL 60510, USA}
\author{A. S. Chou} 
  \affiliation{Fermi National Accelerator Laboratory, Batavia IL 60510, USA}
  \author{M. Hollister} 
  \affiliation{Fermi National Accelerator Laboratory, Batavia IL 60510, USA}
    \author{S. Knirck} 
  \affiliation{Fermi National Accelerator Laboratory, Batavia IL 60510, USA}
\author{A. Sonnenschein} 
  \affiliation{Fermi National Accelerator Laboratory, Batavia IL 60510, USA}
  \author{W. Wester} 
  \affiliation{Fermi National Accelerator Laboratory, Batavia IL 60510, USA}
    \author{R. Khatiwada}
     \affiliation{Illinois Institute of Technology, Chicago IL 60616, USA}
     \affiliation{Fermi National Accelerator Laboratory, Batavia IL 60510, USA}

\author{J. Brodsky}
\affiliation{Lawrence Livermore National Laboratory, Livermore, CA 94550, USA}
\author{G. Carosi}
\affiliation{Lawrence Livermore National Laboratory, Livermore, CA 94550, USA}

\author{L. D. Duffy}
  \affiliation{Los Alamos National Laboratory, Los Alamos, NM 87545, USA}

\author{M. Goryachev}
\affiliation{University of Western Australia, WA, Australia}
\author{B. McAllister}
\affiliation{University of Western Australia, WA, Australia}
\author{A. Quiskamp}
\affiliation{University of Western Australia, WA, Australia}
\author{C. Thomson}
\affiliation{University of Western Australia, WA, Australia}
\author{M. E. Tobar}
\affiliation{University of Western Australia, WA, Australia}

\author{C. Boutan}
  \affiliation{Pacific Northwest National Laboratory, Richland, WA 99354, USA}
\author{M. Jones}
  \affiliation{Pacific Northwest National Laboratory, Richland, WA 99354, USA}
\author{B. H. LaRoque}
  \affiliation{Pacific Northwest National Laboratory, Richland, WA 99354, USA}
\author{E.~Lentz}
  \affiliation{Pacific Northwest National Laboratory, Richland, WA 99354, USA} 
  \author{N.E. Man} 
  \email[Correspondence to: ]{mguzz28@uw.edu and nicole.man@pnnl.gov}
  \affiliation{Pacific Northwest National Laboratory, Richland, WA 99354, USA}
\author{N. S.~Oblath}
  \affiliation{Pacific Northwest National Laboratory, Richland, WA 99354, USA}
\author{M.~S. Taubman}
  \affiliation{Pacific Northwest National Laboratory, Richland, WA 99354, USA}
\author{J.~Yang}%
  \affiliation{Pacific Northwest National Laboratory, Richland, WA 99354, USA}

\author{John Clarke}
  \affiliation{University of California, Berkeley, CA 94720, USA}
\author{I. Siddiqi}
  \affiliation{University of California, Berkeley, CA 94720, USA}

\author{A. Agrawal}
\affiliation{University of Chicago, IL 60637, USA}
\author{A. V. Dixit}
\affiliation{University of Chicago, IL 60637, USA}
 
\author{J.~R.~Gleason}
  \affiliation{University of Florida, Gainesville, FL 32611, USA}
  \author{Y. Han}
  \affiliation{University of Florida, Gainesville, FL 32611, USA}
  \author{A. T. Hipp}
  \affiliation{University of Florida, Gainesville, FL 32611, USA}
\author{S. Jois}
  \affiliation{University of Florida, Gainesville, FL 32611, USA}
 \author{P. Sikivie}
  \affiliation{University of Florida, Gainesville, FL 32611, USA}

\author{N. S. Sullivan}
  \affiliation{University of Florida, Gainesville, FL 32611, USA}
\author{D. B. Tanner}
  \affiliation{University of Florida, Gainesville, FL 32611, USA}

\author{E. J. Daw}
  \affiliation{University of Sheffield, Sheffield, UK}
  \author{M. G. Perry}
  \affiliation{University of Sheffield, Sheffield, UK}

\author{J. H. Buckley}
  \affiliation{Washington University, St. Louis, MO 63130, USA}
  \author{C. Gaikwad}
  \affiliation{Washington University, St. Louis, MO 63130, USA}
\author{J. Hoffman}
  \affiliation{Washington University, St. Louis, MO 63130, USA}
\author{K. W. Murch}
  \affiliation{Washington University, St. Louis, MO 63130, USA}
  \author{J. Russell}
  \affiliation{Washington University, St. Louis, MO 63130, USA}
 
\collaboration{ADMX Collaboration}\noaffiliation

\date{\today}


\begin{abstract}

The Axion Dark Matter eXperiment (ADMX) has previously excluded Dine-Fischler-Srednicki-Zhitnisky (DFSZ) axions between 680-790 MHz under the assumption that the dark matter is described by the isothermal halo model. However, the precise nature of the velocity distribution of dark matter is still unknown, and alternative models have been proposed. We report the results of a non-virialized axion search over the mass range 2.81--3.31 $\mu$eV, corresponding to the frequency range 680--800 MHz. This analysis marks the most sensitive search for non-virialized axions sensitive to Doppler effects in the Milky Way Halo to date. Accounting for frequency shifts due to the detector's motion through the Galaxy, we exclude cold flow relic axions with a velocity dispersion of $\mathcal{O}$($\mathrm{10^{-7}}$)c  with 95\% confidence.
\end{abstract}
\maketitle

\section{Introduction}
Though some clues about the nature of dark matter can be inferred from a variety of astrophysical observations, a definitive explanation for it has yet to be found. Given the elusive nature of dark matter, scientists and researchers alike have explored a variety of candidates to explain a longstanding mystery in particle physics. One such candidate, the axion, is a product of a popular solution to an intractable puzzle known as the Strong $CP$ problem, in which the absence of a neutron electric dipole moment contradicts (or implies fine-tuning of) the $CP$-violation that is implicit in the theory of quantum chromodynamics (QCD) \cite{Peccei1977Sept,weinberg,wilczek}. The axion is therefore a compelling dark matter candidate because of its properties of electric neutrality and feeble interactions with standard model particles. Additional studies show the axion could account for the entire observed dark matter density \cite{Planck,ABBOTT1983133,DINE1983,PRESKILL1983,Ipser1983}.

The Axion Dark Matter eXperiment (ADMX) is a direct detection experiment searching for axion dark matter. ADMX uses a microwave cavity immersed in a magnetic field, which stimulates the conversion of axions to photons via the Inverse Primakoff effect, a realization of an experimental technique called the axion haloscope \cite{sikiviehaloscope}. The experiment leverages the use of a high quality factor cavity, ultra low-noise amplifiers and a dilution refrigerator to achieve sensitivity to both benchmark models for the QCD axion, namely the Kim-Shifman-Vainshtein-Zakharov (KSVZ)~\cite{KIM1979,SHIFMAN1980} and the Dine-Fischler-Srednicki-Zhitnitsky (DFSZ) models~\cite{DINE1981,Zhitnitsky:1980tq}. The coupling strength in the DFSZ model is particularly compelling due to its compatibility with grand unification theories~\cite{DINE1981}.

ADMX was the first axion haloscope to exclude the DFSZ axion in 2018, covering the mass range 2.66-2.81$\mu$ eV~\cite{Du_2018}. Since then, the experiment has been searching for higher mass axions at equivalent sensitivity~\cite{Run1B_Full,Run1C_P1}. 
 
\subsection{Motivation}
During the course of data taking, ADMX measures power spectra at different cavity frequencies to find narrow power excesses that could be potential axion candidates. ADMX acquires two parallel data sets: a medium resolution (MR) data set and a high resolution (HR) data set. The HR data set is in the form of time series data with a spectral resolution of 10 mHz while the MR data set consists of Fourier transformed power spectra with spectral resolution of 100 Hz. The root-mean-square (RMS) frequency variation of a signal, $\delta f$, can be determined using Eq.~\ref{eqn:spectral_linewidth}, where $f$ is the signal frequency, $E_a$ is the average axion energy, $v$ is the magnitude of the velocity, $\delta v$ is the velocity dispersion and $c$ is the speed of light. The observed spectral linewidth is defined as $\Delta f = 2\delta f$. 

\begin{equation}
    \frac{\delta f}{f} = \frac{\delta E_a}{E_a} \approx \frac{v\delta v}{c^2}.
    \label{eqn:spectral_linewidth}
\end{equation}

The MR acquisition channel searches for virialized axions, corresponding to the isothermal sphere model for the dark matter halo distribution in our Galaxy. Axions of this nature are expected to follow a Maxwell-Boltzmann distribution with a velocity dispersion of $\mathcal{O}$($\mathrm{10^{-3}}$)c. Axions with $m_{a} \approx$ \SI{3.1}{\micro\electronvolt} and  $v \approx$ 220 km/s, would have a spectral linewidth $\mathcal{O}$(1) kHz, making them detectable using the MR data set. Previous searches within ADMX have used this data set to exclude dark matter axions that abide by the isothermal halo model within mass range
 \SI{2.66}-\SI{3.31}{\micro\electronvolt} \cite{Du_2018, Run1B_Full}. Searches for axions that follow alternative models have been proposed, which we explore further in this work.

Astrophysical observations have shown evidence for the caustic ring model, suggesting that local dark matter within our solar system exists as discrete flows \cite{PhysRevD.78.063508, duffy_2006, Duffy_2005}.
This is further supported by cosmological simulations that favor a non-Maxwellian model, where the local density of dark matter exists as a superposition of bound clumps and tidal flows, also sometimes referred to as ``streams" \cite{Stiff2003}. These flows are described as the late infall of dark matter particles that have not had sufficient time to be fully thermalized \cite{duffy_2006, Hoskins_2016}. A theorized consequence of such flows are the formation of caustics, or high dark matter density regions, created by the repetitive turnaround process of infalling dark matter particles that pass the Galactic center each time\cite{Sikivie_1998}. The turn-around points of closest approach are known as inner caustic rings, while those that are most distant are outer caustic spheres \cite{Sikivie1999}. 

It has been predicted from observations that the Earth is located near the caustic ring generated by particles falling in and out of the Milky Way for the fifth time \cite{CHAKRABARTY2021100838}.
Depending on whether the Earth sits inside or outside this ring's cross-section, it is dominated by either 4 or 2 flows, respectively. The flows are referred to as ``Big'', ``Little'', ``Up'', and ``Down'', where only the Big and Little flows are present if the Earth sits outside the ring's cross-section.

Given that the velocity dispersion of non-virialized axions from flows within the fifth caustic are significantly narrower than that of virialized axions ($\mathcal{O}(10^{-7})c$)~\cite{sikivie2003evidence,PhysRevD.71.043516,PhysRevLett.92.111301}, the signal from such flows would appear as a narrow peak ($\mathcal{O}(500~\mathrm{mHz})$) in the power spectrum of a Sikivie-type detector like ADMX~\cite{sikiviehaloscope}.

These spectral linewidths are much smaller than the MR bin width of 100 Hz, meaning that all information about spectral shape for such signals is lost in the MR channel. Additionally, the signal-to-noise ratio is significantly reduced because only a fraction of a bin width's worth of signal power is competing with a full bin width's worth of noise power.

Therefore, the HR channel can be used to our advantage. With a bin width of 10 mHz, signals with small velocity dispersions deposit a larger fraction of their power in a single bin than they would in the MR channel, thus achieving a higher signal-to-noise ratio. Additionally, the ADMX detector's relative velocity with respect to an axion causes a Doppler shift in the detected frequency due to the Earth's motion in the Galaxy. To quantify this shift we can look at the total energy of non-relativistic relic axions, ${E_{a}}$, 

\begin{equation}
    E_{a} = m_{a}c^{2} + \frac{1}{2}m_{a}(\overrightarrow{v} \cdot \overrightarrow{v}).
    \label{eqn:axion_energy}
\end{equation} 

Here, $m_{a}$ is the mass of the relic axion, $c$ is the speed of light in a vacuum, and $\overrightarrow{v} = \overrightarrow{v_{a}} -  \overrightarrow{v_{D}}$ is the relative velocity of axion flow and detector. As $\overrightarrow{v}$ changes throughout data-taking, the energy changes slightly, manifesting as a frequency shift of a few tens of Hz at 1 GHz. The MR analysis does not have the required spectral resolution to observe such effects, whereas the HR analysis is capable of detecting these small frequency shifts. The Doppler variation of a candidate signal could therefore be used as a way to quantify its persistence, which we explore further in Section~\ref{persistence} and Appendix~\ref{sec:dopplershift}. 

This paper's focus is on the analysis of the HR data set acquired between 680–800 MHz~\cite{Ram_Thesis, Run1B_Full}. The structure is as follows:  Section~\ref{ADMX Overview} describes the experimental configuration for the data collection for this analysis, with details of the detector and receiver chain, Section~\ref{data acquistion} explains the data acquisition process and details of signal processing,  Section~\ref{cuts} describes the potential candidate identification and examination process, Section~\ref{persistence} and Appendix~\ref{sec:dopplershift} undertake a discussion of the signal modulation and the resulting affects to observed signals, Section~\ref{sec:RFI_check} describes the procedure for distinguishing between radio frequency interference (RFI) and axion-like signals, and Section~\ref{final} provides the limit-setting procedure and interpretation. Barring the existence of any persistent candidates, the limit setting process marks the final step in the HR data analysis, resulting in a statement of exclusion over the 680-800 MHz frequency range for non-virialized axions.

\section{ADMX Overview} \label{ADMX Overview}
\subsection{Detector}
The Axion Dark Matter eXperiment is an example of an axion haloscope: an experimental technique first proposed by Pierre Sikivie in 1983 as a means of detecting the so-called ``invisible'' axion~\cite{sikiviehaloscope}. The detector consists of a high quality factor, cylindrical resonant microwave cavity embedded within a high-field, superconducting solenoid magnet. The strong magnetic field converts passing axions into microwave photons of frequency $E_{a}/h$, where $E_{a}$ is the energy of an axion defined in Eq.~\ref{eqn:axion_energy}, and $h$ is the Planck constant. 

The resonator used for the frequency range of 680–800 MHz is a 140-liter copper-plated stainless steel cavity \cite{Run1B_Full}. The resultant power of the converted photons is amplified when their frequency matches the resonant frequency of the cavity. To leverage resonant enhancement across a range of possible axion masses, ADMX uses two copper tuning rods inside the cavity that can be rotated to alter cavity geometry and its resonant frequency. 

\begin{figure}[h!]
  \centering
    \includegraphics[width=0.4\textwidth]{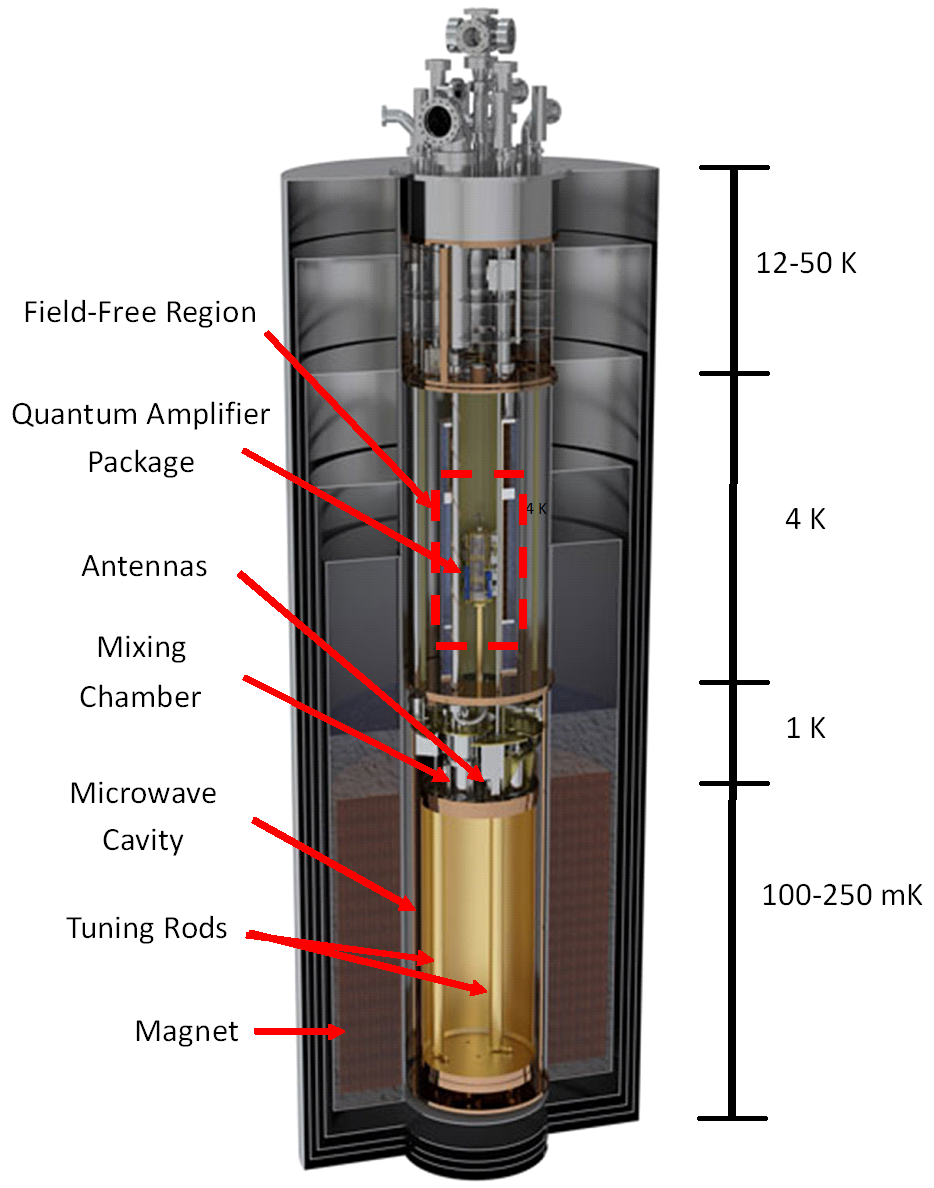}
    \caption{Cutaway diagram of the ADMX insert. The scale on the right hand side represents the different temperature stages of the experiment, with the coldest stage containing the microwave cavity and the first stage cold electronics, such as the Josephson Parametric Amplifier (JPA), as well as the dilution refrigerator. It is worth noting that the JPA is physically located inside the ``Field-Free Region'' because it is highly sensitive to magnetic fields, but it is thermally sunk to the coldest temperature stage. The JPA is also encased in a mu-metal shield meant to provide passive protection from the main magnetic field.}
    \label{fig:insert1}
\end{figure}

A diagram of the ADMX ``insert'', which contains the various cryogenic components of the experiment such as the cavity, cold electronics, and dilution refrigerator, can be seen in Fig.~\ref{fig:insert1}. 

The power is extracted from the cavity using a dipole antenna. An axion signal would manifest as a small narrow-band excess in the resulting digitized power spectrum at a particular frequency corresponding to the axion mass. The expected axion signal power under these experimental conditions is given by Eq.~\ref{Eq:Axion_Power},

\begin{equation}
\begin{split}
    P_{\text{axion}}=2.2{\times}10^{-23}~\mathrm{W}\left(\frac{\beta}{1+\beta}\right)\left(\frac{V}{136~\mathrm{\ell}}\right)\left(\frac{B}{7.6~\mathrm{T}}\right)^2 \\
    \times\left(\frac{C_{010}}{0.4}\right)\left(\frac{g_{\gamma}}{0.36}\right)^2\left(\frac{\rho_{\mathrm{DM}}}{0.45~\mathrm{GeV cm^{-3}}}\right) \\
    \times\left(\frac{f_{a}}{740~\mathrm{MHz}}\right)\left(\frac{Q_{\text{L}}}{30{,}000}\right)\left(\frac{1}{1+(2{\delta}{f_a}/{{\Delta}f_{0}})^2}\right) .
\label{Eq:Axion_Power}
\end{split}
\end{equation}
In the equation above, $V$ is the cavity volume, $B$ is the external magnetic field strength, $C_{010}$ is the form factor (amount of overlap between the electric field of the cavity resonant mode and the external magnetic field), $g_{\gamma}$ is the model-dependent coupling term, $\rho_{\mathrm{DM}}$ is the local dark matter density, $f$ is the frequency of the observed photon, and $Q_L$ is the loaded quality factor of the cavity. The ADMX collaboration assumes a local dark matter density of $\rho_{DM} = \rho_{a} = $0.45 GeVcm$^{-3}$~\cite{Read2014} in presenting its sensitivity to axions that follow the isothermal halo model, hence why the density term is normalized using this value. The cavity coupling parameter, $\beta$, is a measure of how much power is picked up by the strongly coupled antenna. It is defined as $\beta = Q_0/Q_L -1$, where $Q_0$ is the unloaded quality factor. The cavity mode linewidth, $\Delta f_0$, is defined by $\Delta f_0 = f_0/Q_L$. Finally, $\delta f_a$ is some offset from the cavity's resonant frequency, known as the detuning factor. 

ADMX optimizes the magnetic field strength, cavity volume, quality factor, form factor and cavity coupling parameters, in order to maximize the power of an axion signal. In addition, an ultra-low noise amplifier and dilution refrigerator are used to minimize system noise temperature within the experiment itself. These work together to keep the system noise low, increasing the overall signal-to-noise-ratio (SNR). A more detailed description of the detector can be found in Ref. \cite{Run1B_Full, Khatiwada2020AxionDM}.

\subsection{Receiver Chain}
\begin{figure}
    \centering
    \includegraphics[width=0.4\textwidth]{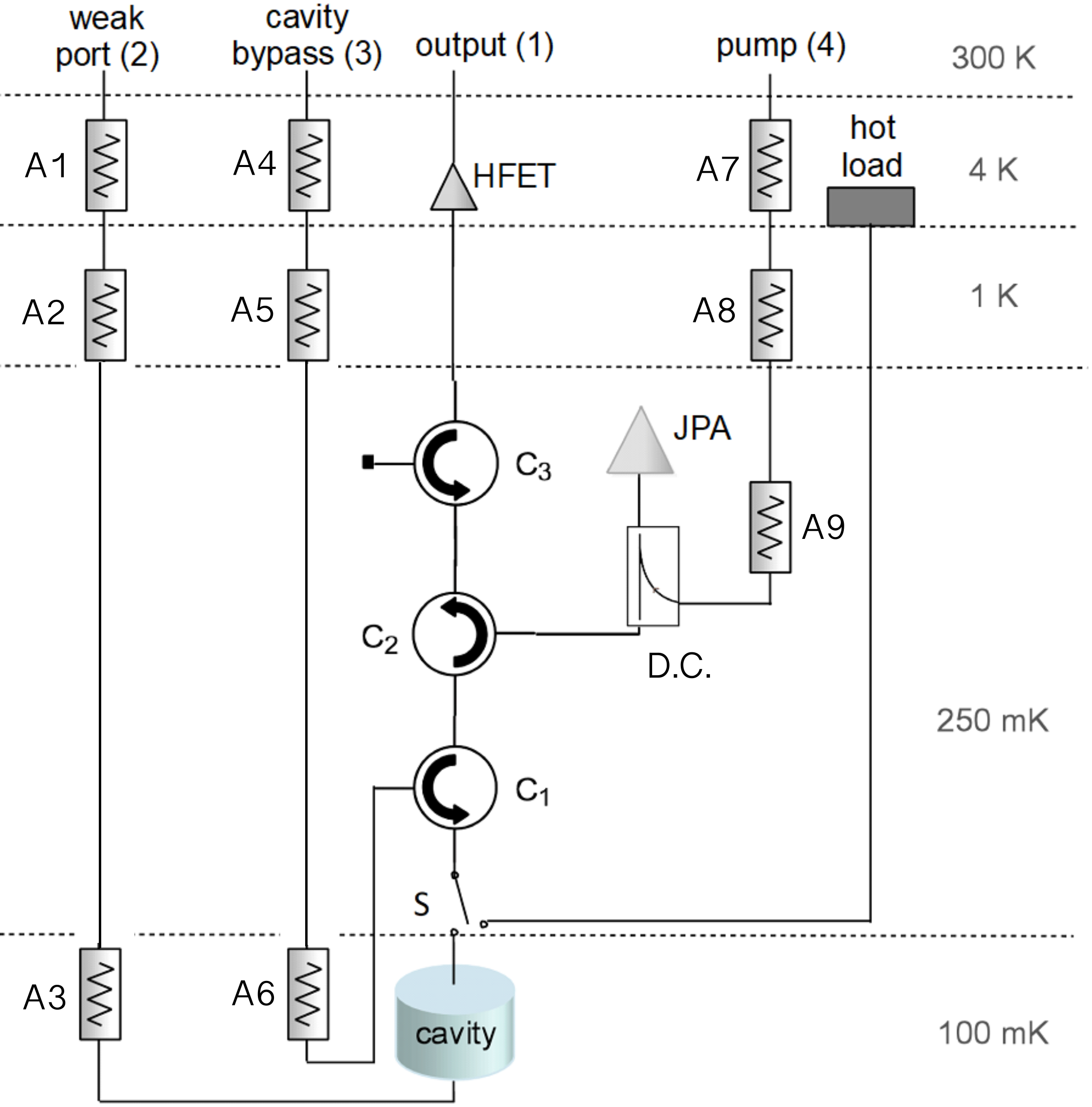}
   \caption{ADMX cold receiver chain for this data-taking run. Components labeled $C_{N}$ are circulators. Components labeled $A_{N}$ are attenuators which are used to thermally sink the RF lines as well as to prevent signals from traveling ``downstream'' toward the cavity. Temperature stages for different components are indicated on the right hand side. The weak port and cavity bypass lines are used for RF calibration measurements, the output line is used to readout power from the cavity, and the pump line supplies the pump tone for the JPA.}
    \label{fig:run1b_receiver_chain}
\end{figure}

The receiver chain extracts and amplifies the signal from the cavity and transmits it to a digitizer where the time series data are both recorded as is (for HR analysis) or Fourier transformed and saved as power spectra (for MR analysis). Additionally, the receiver chain allows for the characterization of various experimental parameters. 
Broadly speaking, the receiver chain is divided into ``warm'' (room temperature) and ``cold'' (cryogenic) components. The cold receiver chain used during this data-taking run can be seen in Fig.~\ref{fig:run1b_receiver_chain}. It consists of a number of lines to characterize the state of the experiment, drive the amplifiers, and readout signal from the cavity.

The weak port (2) as depicted in Fig.~\ref{fig:run1b_receiver_chain}, is designed to perform transmission measurements to determine the cavity's resonant frequency as well as the loaded quality factor, $Q_L$. The cavity bypass line (3) can be used to perform reflection measurements, which allow us to determine how well coupled the antenna is to the mode of the cavity. Details on how these measurements are used to extract these parameters can be found in~\cite{bartram2021axion}. The output line (1) is used to amplify and measure signals coming from the cavity. Signals along this line are first amplified by a Josephson Parametric Amplifier (JPA)~\cite{JPA1,JPA2}, before being amplified further by a Heterostructure Field-Effect Transistor (HFET) from Low Noise Factory~\cite{LNF}. There are also three circulators on the output line which are unidirectional devices used to control the direction of signal flow (as indicated by the arrows) and prevent the amplified signal from the JPA from being reflected back into the cavity. Lastly, there is the pump line (4), which supplies the pump tone for the JPA, allowing us to adjust the JPA's resonant frequency. 

Further information about the warm receiver chain and system noise calibration for this data taking run can be found in Ref.~\cite{bartram2021axion}. 

\section{Data Acquisition} \label{data acquistion}
The ADMX collaboration acquired data from January to October of 2018 that spanned a frequency range 680-800 MHz. The HR data set consisted of a total of 143,249 power spectra. Before Fourier transforming them, the data is stored in time-series form, allowing for flexibility in the choice of spectral bin width. 

Each 100-s digitization has an accompanying sequence of measurements and procedures needed to characterize and optimize the receiver chain, as described in detail in Ref.~\cite{bartram2021axion}. These RF characterization measurements are stored in a database with an associated timestamp that can be mapped to individual time-series data files, along with temperature sensor data. The form factors, $C_{010}$, are simulated using CST Studio~\cite{CST}. Simulation outputs form factors at a few, select frequencies; thus, form factors at every point in frequency space are obtained by interpolating the simulated data.

Each pass through this 100-s sequence is referred to as a single data-taking cadence. Under ideal operating conditions, this data-taking cadence continued, with the tuning rods rotating at a uniform speed in one direction, for approximately 10 MHz, after which point the rods are turned around to rescan. Once the desired sensitivity is achieved, the same region is rescanned at a significantly increased tuning rate, only slowing down at the axion candidate frequencies identified in the MR channel. 

\subsection{Signal Processing} 
\begin{figure}[h!]
    \centering
\includegraphics[width=0.48\textwidth]{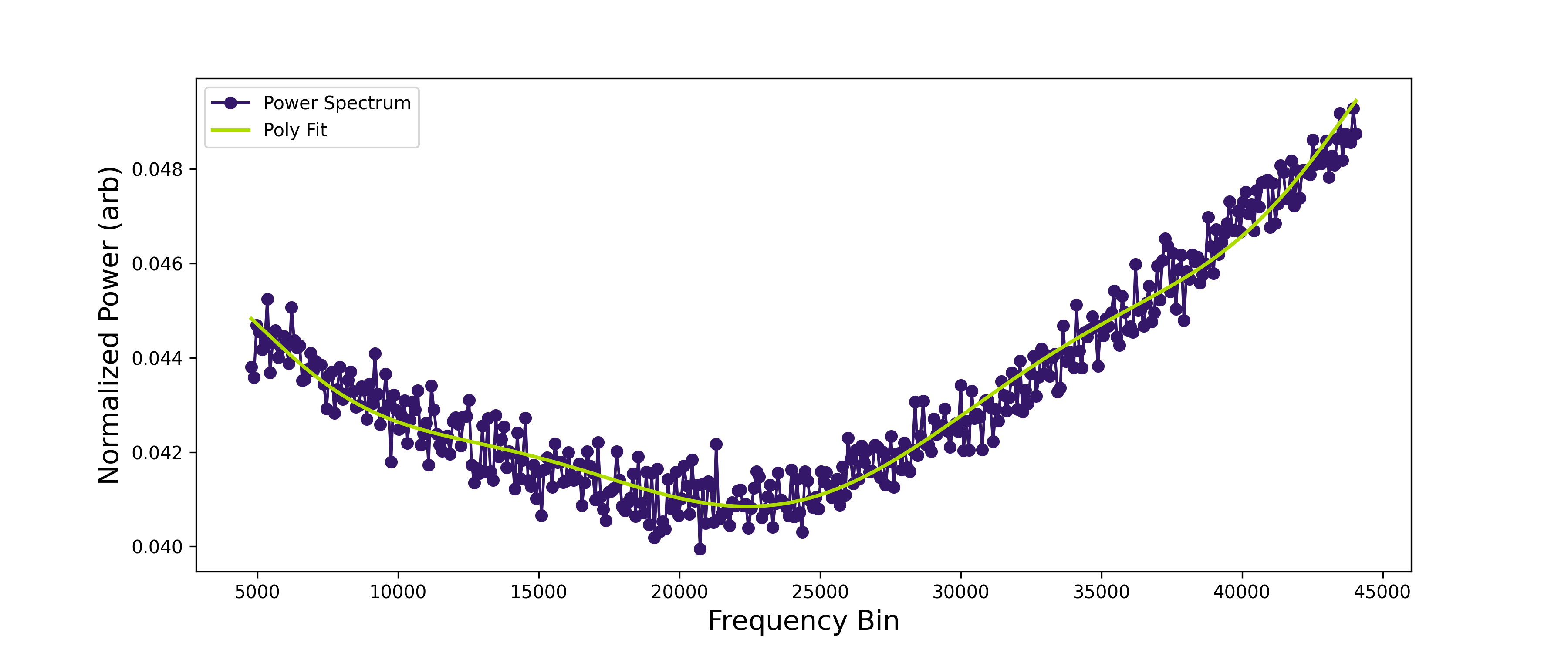}
    \caption{Example power spectrum showing the 8th order polynomial fit to the data. The data has been rebinned such that the plot shows only 512 data points. The HR spectra contains millions of bins -- too many to show here.}
    \label{fig:timeseries}
\end{figure}

Data for the HR channel are originally acquired in the time domain and must therefore be transformed into the frequency domain to perform a search. A sample of the raw data acquired in the frequency domain can be seen in Fig.~\ref{fig:timeseries}. 

The receiver chain imposes a shape on the raw data that changes with frequency. The shape becomes evident once the time-series data are Fourier transformed into power spectra. Removing this shape is the first step in processing the power spectra. For this analysis, we used an 8th order polynomial fit. The fit was then divided out of the data both removing the receiver shape and normalizing it into units of RMS noise power, $\sigma = k_BbT_n$, where $k_B$ is the Boltzmann constant, $b$ is the bin width of the spectrum, and $T_n$ is the system noise temperature.   

\section{Candidate Identification and Examination} \label{cuts}
Potential axion candidates in the power spectrum are identified as frequency bins in which the power exceeded a threshold, $P_{T}$, and are flagged as triggers in the analysis. As outlined in Refs.~\cite{Leanne_2006,haystac_analysis}, the functional form of the noise power distribution for a single bin is given by a $\chi^2$ distribution of degree 2:  

\begin{equation}
    \frac{dP}{dp} = \frac{1}{\sigma}e^{-(\frac{p}{\sigma})} .
    \label{eqn:single_bin_noise}
\end{equation} 

\begin{figure}[H]
    \centering
\includegraphics[width=0.48\textwidth]{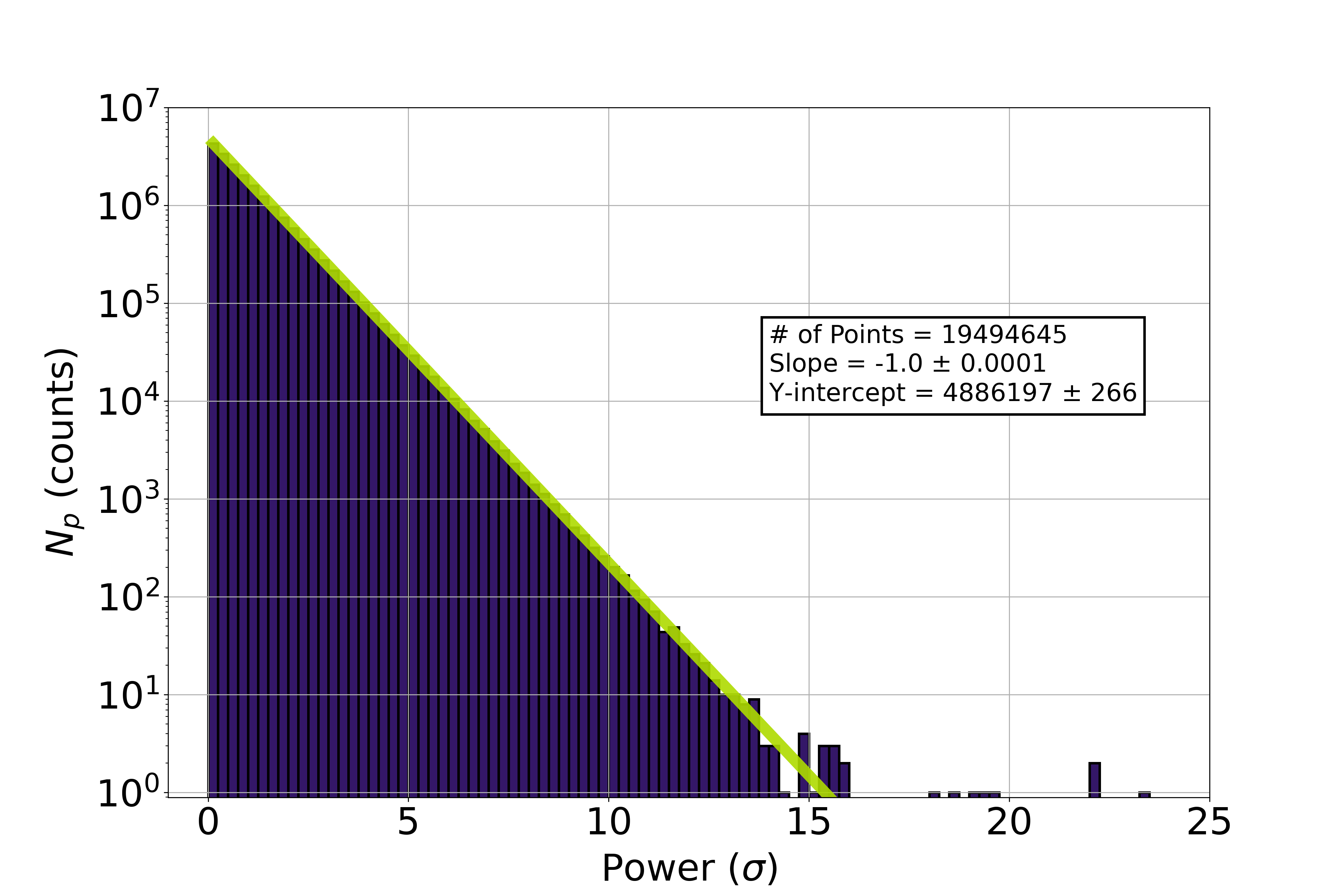}
    \caption{Histogram of 5 high-resolution power spectra. The noise in the HR channel follows an exponential distribution. Power is expressed in terms of $\sigma = k_{B}bT_{n}$, the RMS noise power of a single bin, where $k_{B}$ is the Boltzmann constant, b is the spectral resolution (10 mHz), and $T_{n}$ is the system noise temperature \cite{Leanne_2006}. Given that the energy distribution is proportional to a Boltzmann factor ($e^{-E/k_BT}$) with energy (E) proportional to the square of the amplitude (e.g. $E = mv^2/2$), the noise amplitude for a single component has a Gaussian probability distribution. The addition of these two Gaussian components for a single-bin analysis results in a $\chi^2$ distribution of degree 2 (a simple exponential) as displayed above.}
    \label{fig:hist_noise}
\end{figure}

In order to determine $P_{T}$, we plot a histogram of the normalized power, $p$, using 5 individual power spectra to determine where the tail of the noise distribution lands (see Fig. \ref{fig:hist_noise}). We use more than just one spectrum in order to introduce some level of averaging over fluctuations in a single spectrum. Due to the large number of data points per spectrum however, we chose to use only 5 spectra to keep the computation time reasonable.

According to Eq. \ref{eqn:single_bin_noise}, the number of frequency bins, $N_{p}$, with normalized power between $p$ and $p + \Delta p$ is 

\begin{equation}
    N_{p} = \frac{N \Delta p}{\sigma}e^{-(\frac{p}{\sigma})} ,
    \label{eqn:full_noise_dist}
\end{equation} 

\noindent where $N$ is the total number of points and $\Delta p =$ 0.25 $ \sigma $ is the bin width of the histogram. From Fig. \ref{fig:hist_noise} we can see that the data fit the expected noise distribution well, with mean~=~$\sigma=$~1 (due to power normalization). Additionally, the y-intercept, which corresponds to $N_{p}$ when $p =$~0, follows Eq. \ref{eqn:full_noise_dist} well. With $N$ = 19494645, $\Delta p =$ 0.25 $ \sigma $, and $p =$~0, $N_{p,\mathrm{calc}} =$ 4873661. Dividing this value by the fitted y-intercept of 4886197 gives $N_{p,\mathrm{calc}}/N_{p,\mathrm{fit}}$ = 0.997, indicating that these two values are in quite good agreement. 

Now that we have confidence in our understanding of the noise distribution, we are able to identify a reasonable power threshold. We want to set the threshold such that we can expect to see, on average, 1 noise peak per spectrum that surpasses $P_{T}$, thus eliminating the majority of the noise while maintaining potential signals. To determine this, we calculate the indefinite integral of the fit of the histogram in Fig. \ref{fig:hist_noise}, set it equal to 5 (number of spectra used), and solve for power. This calculation results in $P_{T}$ = 13.8 $\sigma$. To keep things simple, we opted to round this value to the nearest integer, settling on $P_{T}$ = 14 $\sigma$ as our final threshold. 

All triggers in a spectrum that meet or exceed this threshold are categorized as potential axion candidates. Applying this threshold selects 444,968 triggers. The triggers are then down-selected by a series of data cleaning cuts on experimental parameters such as the quality factor and system noise, and proximity (in frequency space) to the resonant frequency of the cavity. The results of these cuts are shown in Tab.~\ref{tab:analysis_cuts}. Additionally, persistence and consistency in calculated density of a signal between spectra is used to further narrow down potential axion candidates. The details of these various checks will be covered in the following sections.

\begin{center}
\begin{table}[H]
\centering
\renewcommand{\arraystretch}{1.3} 
\begin{tabular}{ |m{2.4cm}| m{1.5cm} | m{3.6cm} |} 
\toprule[0.1ex]
\hline
\toprule[0.1ex]
Cut Parameter & Triggers \newline Removed & Constraint \\
\hline
Quality Factor & 1,088 & 10,000 \textless\,$Q_L$\,\textless 120,000 \\
\hline
 System Noise & 2,076 & 0.1 K
 \textless\,$T_{\text{sys}}$\,\textless 2 K \\ 
 \hline
 SAG & 25,735 & N/A  
 \\
 \hline
 \bottomrule[0.1ex]
 \hline
\end{tabular}
\caption{Table of cuts made on potential axion candidates (triggers). The ``triggers removed'' here represent the number of triggers that failed each cut individually. Therefore, there is some overlap between each of these cuts, i.e. some triggers that failed one cut may have (and likely did) fail another cut as well.}
\label{tab:analysis_cuts}
\end{table}
\end{center}

\subsection{Quality Factor}
\label{sec:QFactor}
The quality factor cut is made to omit peaks associated with power spectra which have quality factors greater than $Q_L=$ 120,000 and less than $Q_L=$ 10,000. Power spectra with quality factors outside of this range are likely the result of a poor Lorentzian fit to the transmission measurement. In particular, quality factors greater than $Q_L=$ 120,000 are unphysical given the copper-plating of the ADMX cavity. The quality factor typically varies throughout the run since it is a function of temperature, frequency, and the cavity coupling ($\beta$), however it should be within the range of 10,000 to 120,000 under normal operating conditions. 

\subsection{System Noise Temperature}
\label{sec:Tsys}

The total system noise temperature can be characterized by 
\begin{equation}
T_{\mathrm{sys}} = \frac{T_{H}}{\epsilon\mathrm{SNRI}},
\end{equation}
where $T_{H}$ is the noise temperature of the HFET that is used in ADMX operations as a second-stage amplifier, $\epsilon$ is a measure of signal attenuation between the cavity and output, and SNRI is the signal-to-noise-improvement. This number was found to be between $\sim$11-22 K across the observed frequency range via a standard y-factor measurement\cite{bartram2021axion}. SNRI is the JPA signal-to-noise-improvement-ratio defined as:
\begin{equation}
\mathrm{SNRI} = \frac{G_{\mathrm{on}}}{G_{\mathrm{off}}}\frac{P_{\mathrm{off}}}{P_{\mathrm{on}}}.
\label{eqn:SNRI}
\end{equation}
The on and off subscripts refer to the JPA pump tone being on, or off, respectively, with $G$ being the receiver gain, and $P$ the output power measured under both conditions. When the pump tone is on, the JPA behaves as an amplifier, and when the pump tone is off the JPA behaves as a perfect reflector. 

All triggers with a system noise temperature outside of the range $T_{\mathrm{sys}} = 0.1 - 2$ K are cut from our candidates list. System noises greater than $T_{\mathrm{sys}} = 2$ K are likely due to anomalous JPA SNRI measurements. System noise temperatures less than $T_{\mathrm{sys}} = 0.1$ K would be lower than the standard quantum limit and, therefore, unphysical. Measurements of such low system noise temperatures can arise from overestimation of the JPA SNRI, typically due to a poor gain measurement. 

\subsection{Full Width at Half Maximum}
\label{sec:FWHM}

Triggers at frequencies outside of the Full Width at Half Maximum (FWHM) of the cavity Lorentzian are removed because an axion signal at a frequency outside of this region would have its power reduced by at least one-half its original strength. The frequency range included in the FWHM is defined by $f_{0} \pm \frac{f_{0}}{2Q_L}.$ In this equation $f_{0}$ is the resonant frequency of the cavity mode, and $Q_L$ is the loaded quality factor. In our analysis we opt to use a slightly more conservative range defined by $f_{0} \pm \frac{f_{0}}{1.8Q_L}.$ This preserves more data such that we can more easily evaluate whether the signal originated from within or from outside the cavity later on in the analysis. 

\subsection{Synthetic Axion Removal}
\begin{figure}[h]
  \centering
    \includegraphics[width=0.48\textwidth]{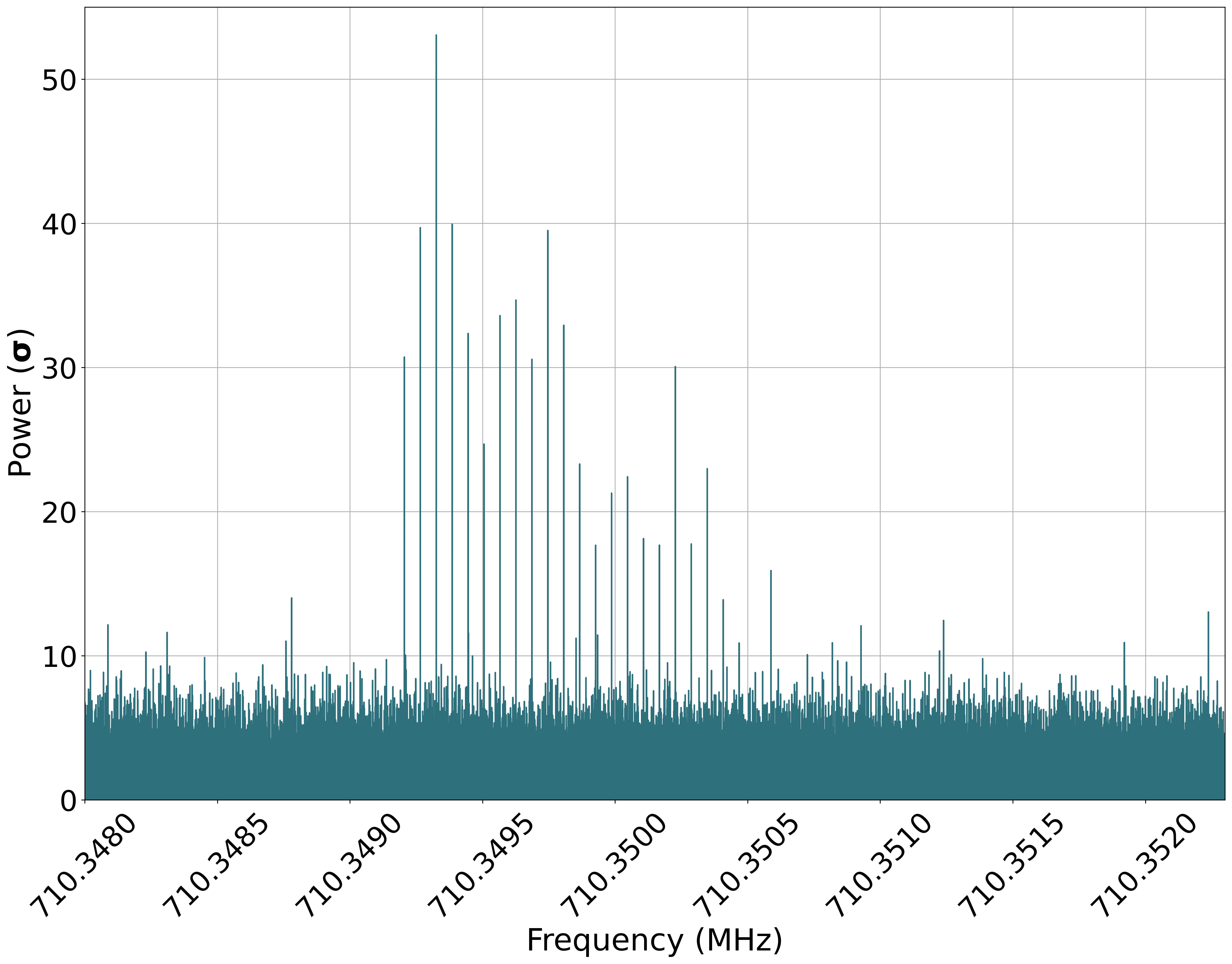}
    \caption{High resolution spectrum containing a synthetically generated axion signal which is used for the medium resolution analysis. The ability to see the evenly spaced bins with particularly high powers make it clear that this is a synthetic signal rather than a true axion signal, thus we can safely remove this for this data analysis.}
    \label{fig:SAG}
\end{figure}

The ADMX medium-resolution analysis procedure involves blind injection of synthetic axion signals into the cavity. These are used to ensure our analysis is capable of detecting an axion signal, were it to appear. However, these synthetic signals are binned in such a way that it is obvious they are not real when looking at the HR data, making them useless in terms of testing this analysis. Thus, for the purposes of this paper, they are simply identified and removed. An example of a synthetic signal in a high resolution spectrum can be seen in Fig.~\ref{fig:SAG}. 

\subsection{Persistence} \label{persistence}
The axion signal frequency as seen from an Earth-based haloscope experiences both diurnal and annual modulations due to the Earth's motion with respect to the Galactic halo. The velocity of a stationary detector on the Earth's surface with respect to the rest frame of the Galactic halo  ($\vec{v}_{D}$) is comprised of three terms:
\begin{equation}
\vec{v}_{D}=\vec{v}_{LSR}+\vec{v}_{\odot,LSR}+\vec{v}_{o}+\vec{v}_{r}.
\label{Eq:Doppler_Vel}
\end{equation}
Here,$\vec{v}_{LSR}$ is the velocity of the Local Standard of Rest, $\vec{v}_{\odot,LSR}$ is the Sun's velocity with respect to the LSR, $\vec{v}_{o}$ is the Earth's orbital velocity, and $\vec{v}_{r}$ is the Earth's rotational velocity at the location of the detector. 

The frequency of an axion signal in ADMX is defined by $f_a = E_a/h$, where $E_a$ is the axion energy as defined in Eq.~\ref{eqn:axion_energy} and $h$ is the Planck constant. Since the axion energy depends on the overall velocity, $\vec{v} = \vec{v}_{a}-\vec{v}_{D}$, and $\vec{v}_{D}$ changes over time, we can expect the frequency of an axion signal to drift slightly over the course of data taking, on scales that are detectable with the high frequency resolution of this data set. Signals that drift more in frequency than is expected from the change in the detector velocity would not be real axion signals. Therefore, a cut can be imposed by considering Doppler frequency shift, which we define as the persistence cut.

We calculate the maximum Doppler shift that could have been observed during the data acquisition for frequency range 680-800 MHz, for five different axion velocity models (details in Appendix~\ref{sec:dopplershift}), and the results are summarized in Tab.~\ref{tab:doppler}.

\begin{center}
\begin{table}[H]
\centering
\renewcommand{\arraystretch}{1.5} 
\begin{tabular}{ |m{4cm}| m{2cm} | } 
\toprule[0.1ex]
\hline
\toprule[0.1ex]
\bf{Name of Flow} & $\boldsymbol{\Delta_{f}}$ \bf{(Hz)}  \\
\hline
Big  & 56 \\
\hline
Little  & 48 \\
\hline
Up  & 49 \\
\hline
Down  & 68 \\
\hline
\bottomrule[0.1ex]
\hline
\end{tabular}
\caption{Maximum observed Doppler shifts for different axion flows. Calculated for $f_0 = 800$ MHz with a $\Delta_t$ of 62 days.}
\label{tab:doppler}
\end{table}
\end{center}

Based on these results, we set the maximum drift frequency to $\pm 68\:\mathrm{Hz}$ as this would safely account for drifts from the four discrete flows that the Earth is expected to be in \cite{CHAKRABARTY2021100838}, as well as the naive radial infall case. Further, because our measure of persistence should take into account the number of times a given frequency is observed, we calculate a quantity known as the persistence ratio, $\gamma$. This ratio measures the fraction of total spectra a given trigger appeared in, and is defined by

\begin{equation} \label{eqn:persistence_ratio}
    \gamma = \frac{\text{\#\:of\:triggers\:within\:drift\:freq.}}{\text{\#\:of\:spectra\:that\:contain\:trigger\:freq.}}.
\end{equation}

\begin{figure}[h!]
  \centering
    \includegraphics[width=0.48\textwidth]{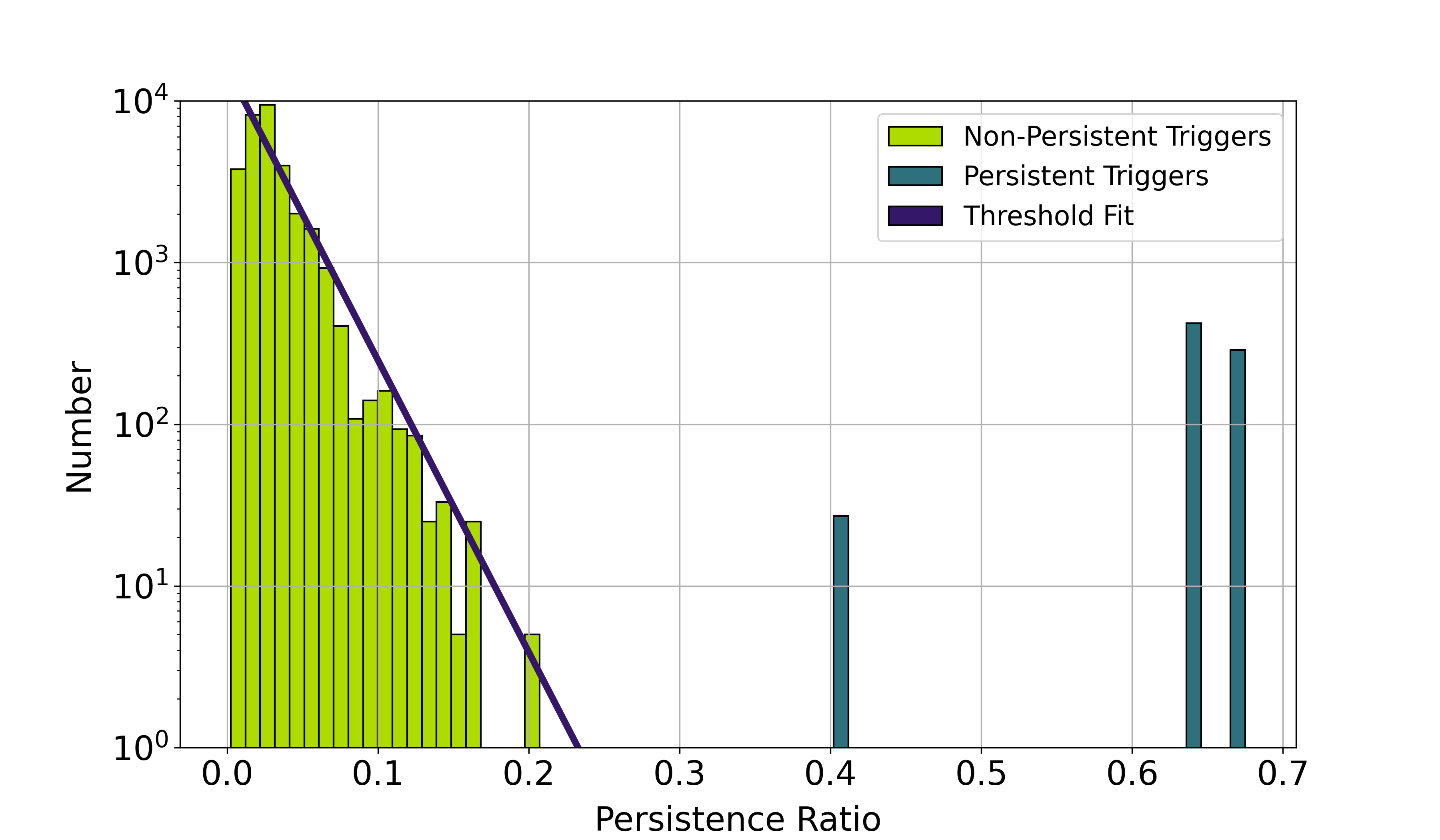}
    \caption{Histogram of persistence ratio for triggers that pass all initial cuts and have at least one other trigger within $\pm68\:\mathrm{Hz}$. An exponential fit of the data is shown in purple. The majority of the triggers (shown in green) fall within the bounds of the fit, however three triggers have particularly large persistence ratios when compared to the rest of the data. We denote these as persistent triggers, shown in teal.}
    \label{fig:persistence_ratio}
\end{figure}

The persistence ratio is only calculated for triggers that pass all the aforementioned cuts and have at least one other trigger within the drift frequency ((numerator of $\gamma)\ge1$). This leaves 34,933 candidates to calculate the persistence ratio of. A distribution of $\gamma$ for these remaining triggers can be seen in Fig.~\ref{fig:persistence_ratio}.

Based on the fit of the distribution, shown in Fig.~\ref{fig:persistence_ratio}, it is obvious that the triggers with persistence ratios with $\gamma >$ 0.2 (highlighted in teal) are outliers from the rest of the data. As these signals are particularly persistent, they warrant further investigation.

\subsection{Signal Origin Test}
\label{sec:RFI_check}
The persistent trigger frequencies are 686.3096, 742.5591, and 792.6526 MHz. To further determine the nature of these triggers, we perform a check to distinguish between axion-like signals, which originate from within the cavity, and RFI from external sources. This is typically done in the MR analysis by examining the power of the signal as the cavity frequency is tuned on and off resonance with the signal frequency. When $f_{0}$ = $f_{a}$, where $f_{0}$ is the cavity frequency and $f_{a}$ is the trigger frequency, the power should be enhanced, and as the cavity tunes away from $f_a$, the power should fall off following a Lorentzian line shape. 

\begin{figure}[h!]
    \centering
\includegraphics[width=0.45\textwidth]{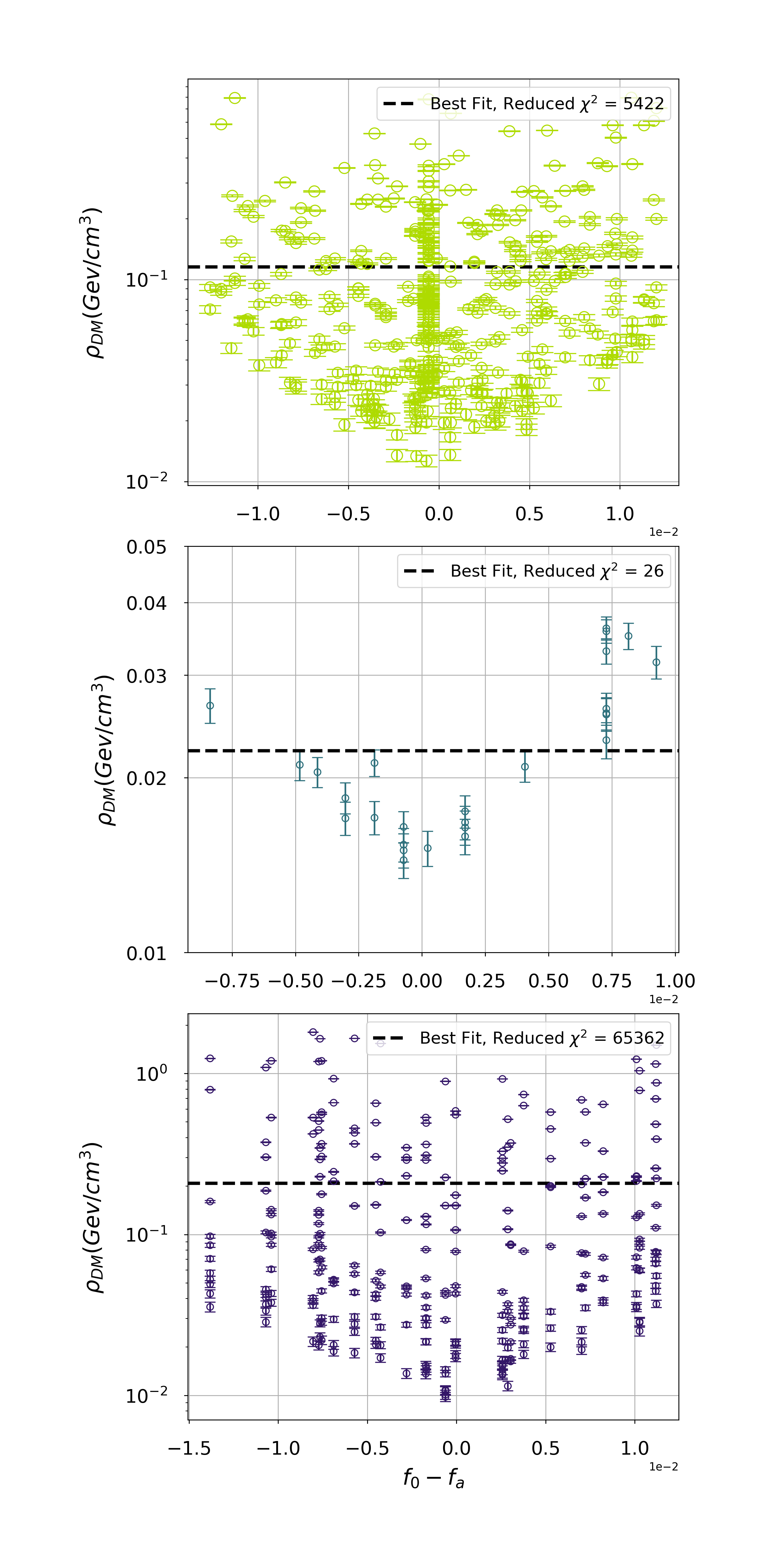}
    \caption{Dark matter density ($\rho_{\mathrm{DM}}$) vs. offset of the trigger frequency from the cavity frequency ($f_{0} - f_{a}$) for the three persistent triggers we found (Top: 686.3096 MHz, Middle: 742.5591 MHz, Bottom: 792.6526 MHz). Dark matter density for a given signal (assuming it comes from a single source) should be constant, thus constant fits of the data are shown. Reduced $\chi^2$ values are shown for each of the three fits in the legend. The number of data points on each plot is related to the number of times that each trigger was observed. Frequencies near the trigger at $\sim$742 MHz were observed much less than the other two, hence why that plot has fewer points.}
    \label{fig:RFI_check}
\end{figure}

For the HR analysis, however, differences in parameters such as quality factor and system noise temperature have a greater effect on observed power (in units of $\sigma$) due to the lack of spectrum averaging. Thus, we opt to solve for axion density, $\rho_{\mathrm{DM}}$, rather than power, which should be consistent for a real axion signal regardless of cavity parameters (resulting in a flat line when plotting $\rho_{\mathrm{DM}}$ vs $f_{0} - f_{a}$). We achieve this by scaling the normalized power in units of $\sigma$ by the system noise to get power back in units of Watts. We can then solve Eq. \ref{Eq:Axion_Power} for $\rho_{\mathrm{DM}}$ and scale the power ($P_{\mathrm{axion}}$) by the appropriate factors, cancelling out effects caused by differences in cavity parameters between individual spectra. The results of this calculation can be seen in Fig. \ref{fig:RFI_check}. The reduced $\chi^2$ values for 686.3096, 742.5591, 792.6526 MHz were 5422, 26, 65362 respectively. As these are all $\gg 1$, we conclude that the persistent triggers do not follow the constant density model, and are therefore not from axions, but are instead RFI.

\section{Exclusion Limit} \label{final}
\begin{figure*}
  \centering
\includegraphics[width=0.8\textwidth]{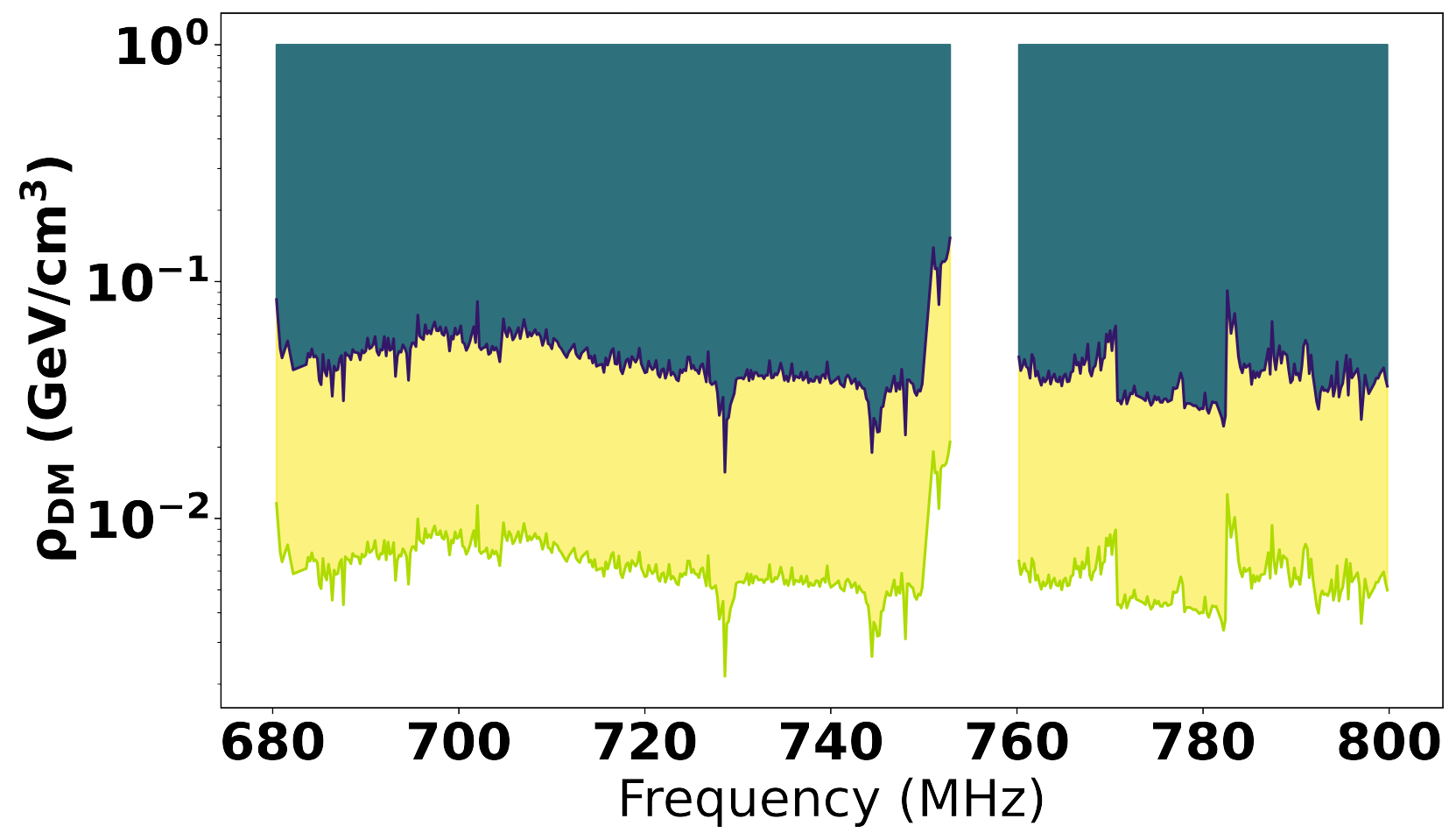}
    \caption{Exclusion limit on the nonvirialized axion dark matter density. Teal shows the exclusion limit assuming DFSZ coupling, whereas yellow shows the exclusion limit assuming KSVZ coupling. The gap between 750 and 760 MHz represents a cavity mode crossing. The Big, Little, Up and Down flows predicted to be inside the cross-section of the fifth caustic ring are expected to have densities of 11.2, 1.1, 5.4, and 4.7 $\mathrm{GeV/cm}^3$ respectively\cite{CHAKRABARTY2021100838}. Thus, axions from all four flows are excluded in this frequency range.}
    \label{fig:exclusion}
\end{figure*}

All candidates in the frequency range of interest are ultimately excluded, enabling us to set a limit. The exclusion limit must account for the tests for persistence, which is unique to the HR analysis. The persistence test requires knowledge of how frequently any given bin is scanned -- a difficult number to retrieve given the sheer number of bins in the HR channel. As a result, we developed a Monte Carlo simulation in which software synthetic axions are generated and injected into the raw time-series data, inspired by the limit setting procedures described in \cite{Daw_thesis,ADMX_2010,MC_limit}. The synthetic axion signal-to-noise ratio was calibrated according to measured RF parameters such as the system noise and quality factor of the detector. The model-dependent coupling term, $g_{\gamma}$, is fixed to the standardized DFSZ (KSVZ) value of 0.36 (0.97). The dark matter density was a free parameter that could be varied over a wide range. We injected a total of 600 synthetic axion signals, with one synthetic signal every 200 kHz, starting at 680 MHz up to 800 MHz. This process was repeated with different values for the dark matter density each time, until 95\% of the software synthetic signals were detected. This value for the dark matter density determines the 95\% confidence level. The resulting exclusion limit can be seen in Fig.~\ref{fig:exclusion}.

\section{Conclusion}
In summary, this paper represents the first HR axion search since the ADMX collaboration started taking data at DFSZ sensitivity. Previously, the MR analysis excluded axions obeying the isothermal halo model in the frequency range 680--790 MHz \cite{Run1B_Full}. Unlike the MR channel, the HR channel is uniquely sensitive to a range of models for the detection of non-virialized axions. Through our analysis, we exclude non-virialized DFSZ (KSVZ) axion signals in the frequency range 680--800 MHz with dark matter densities between $\mathrm{0.02 - 0.09\,GeV/cm^3} \mathrm{(0.003 - 0.01\,GeV/cm^3)}.$

\section{Acknowledgements}
This work was supported by the U.S. Department of Energy through Grants No. DE-SC0009800, No. DESC0009723, No. DE-SC0010296, No. DESC0010280, No. DE-SC0011665, No. DEFG02-97ER41029, No. DEFG02-96ER40956, No. DEAC52-07NA27344, No. DEC03-76SF00098, No. DE-SC-0022148, and No. DESC0017987. Fermilab is a U.S. Department of Energy, Office of Science, HEP User Facility. Fermilab is managed by Fermi Research Alliance, LLC (FRA), acting under Contract No. DE-AC02-07CH11359. Additional support
was provided by the Heising-Simons Foundation and by the Lawrence Livermore National Laboratory and Pacific Northwest National Laboratory LDRD offices. UWA participation is funded by the ARC Centre of Excellence
for Engineered Quantum Systems, Grant
No. CE170100009, Dark Matter Particle Physics, Grant No. CE200100008, and Forrest Research Foundation. The
corresponding author is supported by JSPS Overseas Research Fellowships No. 202060305. LLNL Release No. LLNL-JRNL-853502. Chelsea Bartram acknowledges support from the Panofsky Fellowship at SLAC.

\appendix
\section{Doppler Shift Calculation} \label{sec:dopplershift}
In this Appendix we describe the calculation of the expected Doppler shifts from the four flows of interest predicted by the Caustic Ring Model, following the framework outlined in Ref.~\cite{Turner_1990}. We begin our calculation by determining the individual components of the detector velocity, which we will redefine here for convenience:

\begin{equation}
\vec{v}_{D}=\vec{v}_{LSR}+\vec{v}_{\odot,LSR}+\vec{v}_{o}+\vec{v}_{r}.
\label{Eq:Doppler_Vel_apppendix}
\end{equation}

Here, $\vec{v}_{LSR}$ is the Local Standard of Rest (LSR) velocity, $\vec{v}_{\odot,LSR}$ is Sun's velocity with respect to the LSR, $\vec{v}_{o}$ is the Earth's orbital velocity around the Sun, and $\vec{v}_{r}$ is the Earth's axial rotational velocity at the location of the detector. We will use a coordinate system set in the Galactic Rest Frame ($\hat{r}$, $\hat{\phi}$, $\hat{z_g}$) where $\hat{r}$ points away from the Galactic center, $\hat{\phi}$ points in the direction of the Sun's orbital motion in the Galactic plane, and $\hat{z_g}$ points along the North Galactic Pole. The plane of the Earth's rotation about the Sun (known as the ecliptic) is tilted with respect to the Galactic plane, mostly in the  $\hat{\phi}- \hat{z_g}$ plane, at an angle of about $60^\circ$. 

We assume a value of $\vec{v}_{LSR} = 220 $ km/s $\cdot \; \hat{\phi}$ (up to about 10$\%$ error). 

There are a number of estimates of the solar peculiar velocity, but we opt to use  $(\vec{v}_{\odot,LSR,r},\vec{v}_{\odot,LSR,\phi},\vec{v}_{\odot,LSR,z_g}) = (-8.63 \pm 0.64, \; 4.76 \pm 0.49, \; 7.26 \pm 0.36)$ km/s~\cite{Ding_2019_LSR}.

Using the vernal equinox as the location where t = 0, we can write the Earth's orbital velocity as defined below.

\begin{equation}
\begin{split}
   & \vec{v}_{o,r} = -v_{o}\mathrm{cos}(\omega t)\; \hat{r} \\
   & \vec{v}_{o,\phi} = v_{o}\mathrm{sin}(\omega t)\mathrm{cos}(60^{\circ})\; \hat{\phi} \\
   & \vec{v}_{o,z_g} = -v_{o}\mathrm{sin}(\omega t)\mathrm{sin}(60^{\circ})\; \hat{z_g}
\end{split}
\end{equation}

In these expressions, ${v}_{o}$ and $\omega$ are the linear and angular speeds of Earth's motion around the Sun respectively. We assume a value of ${v}_{o} = 30 $ km/s and $\omega = 2{\times}10^{-7}$ rad/s. 

Earth's rotational speed on the surface at the equator can be calculated as $v_{r,eq} = \omega_s R_\oplus$, where $\omega_s = 7.3{\times}10^{-5}$ rad/s is the angular speed of Earth's rotation and $R_\oplus = 6,378$ km is Earth's radius. To determine what the speed is on the surface at the detector location we must do $v_{r} = v_{r,eq}\mathrm{cos(lat)}$ where $v_{r,eq} = 0.46$ km/s and $\mathrm{lat} = 47^{\circ}$ is the latitude of the detector location (Seattle, WA, USA). 

We can define a parameter, $\varphi$, as the difference between the detector's longitude, and the longitude which is facing the Sun at the vernal equinox measured from west to east. Defining the Earth's rotational velocity in the Galactic frame is non trivial, and it is in fact easier to first define $\vec{v}_{r}$ in a frame centered on Earth, then one centered on the Sun, and finally converting this to the Galactic frame. 

The Earth-centered reference frame has a coordinate system ($\hat{x}, \hat{y}, \hat{z}$), where $\hat{z}$ points in the direction of the Earth's North Pole, $\hat{x}$ points in the direction of the Sun when the Earth is at the vernal equinox, and $\hat{y}$ completes a right-handed coordinate system. 

The Sun-centered reference frame has a coordinate system ($\hat{x'}, \hat{y'}, \hat{z'}$), where $\hat{x'}$ is in the same direction as $\hat{x}$, $\hat{z'}$ is in the direction of the normal to the ecliptic (about $23^{\circ}$ from $\hat{z}$), and $\hat{y'}$ completes a right-handed coordinate system. 

In the Earth-centered frame ($\hat{x}, \hat{y}, \hat{z}$), the components of $\vec{v}_{r}$ can be written as:

\begin{equation}
\begin{split}
   & \vec{v}_{r,x} = -v_{r}\mathrm{sin}(\varphi + \omega_s t)\; \hat{x} \\
   & \vec{v}_{r,y} = v_{r}\mathrm{cos}(\varphi + \omega_s t)\; \hat{y} \\
   & \vec{v}_{r,z} = 0\; \hat{z}.
\end{split}
\end{equation}

Converting these into the Sun-centered frame ($\hat{x'}, \hat{y'}, \hat{z'}$) gives:  

\begin{equation}
\begin{split}
   & \vec{v}_{r,x'} = -v_{r}\mathrm{sin}(\varphi + \omega_s t)\; \hat{x'} \\
   & \vec{v}_{r,y'} = v_{r}\mathrm{cos}(\varphi + \omega_s t)\mathrm{cos}(23^{\circ})\; \hat{y'} \\
   & \vec{v}_{r,z'} = -v_{r}\mathrm{cos}(\varphi + \omega_s t)\mathrm{sin}(23^{\circ})\; \hat{z'}.
\end{split}
\end{equation}

Finally, we can convert these velocities into the Galactic frame ($\hat{r}$, $\hat{\phi}$, $\hat{z_g}$), noting that $\hat{y'}$ is in the same direction as $\hat{r}$. This amounts to a rotation about the $\hat{y'}$/$\hat{r}$ axis of $60^{\circ}$. 

\begin{equation}
\begin{split}
   & \vec{v}_{r,r} =  v_{r}\mathrm{cos}(\varphi + \omega_s t)\mathrm{cos}(23^{\circ})\; \hat{r}\\
    & \vec{v}_{r,\phi} = [-v_{r}\mathrm{sin}(\varphi + \omega_s t)\mathrm{cos}(60^{\circ}
   ) \\ 
   & \qquad \quad -v_{r}\mathrm{cos}(\varphi + \omega_s t)\mathrm{sin}(23^{\circ})\mathrm{sin}(60^{\circ}
   )]\; \hat{\phi} \\
   & \vec{v}_{r,z_g} = [v_{r}\mathrm{sin}(\varphi + \omega_s t)\mathrm{sin}(60^{\circ}
   ) \\ 
   & \qquad \quad -v_{r}\mathrm{cos}(\varphi + \omega_s t)\mathrm{sin}(23^{\circ})\mathrm{cos}(60^{\circ}
   )]\; \hat{z_g}
\end{split}
\end{equation}

We can now calculate the Doppler shift, the equation for which can be derived from Eq.~\ref{eqn:axion_energy}: 
\begin{equation}
    h\Delta f_a =\frac{1}{2}m_{a}\Delta(\vec{v}\cdot\vec{v}).
\end{equation}

In this situation, $\vec{v} = \vec{v}_{a} - \vec{v}_{D}$, where $\vec{v}_{a}$ is the axion velocity, and $\vec{v}_{D}$ is the detector velocity as defined above. As one may notice, this shift is dependent on the axion mass and will increase with increasing mass. Considering this, we used the high end of our mass/frequency range (3.31 $\mu eV$/800 MHz) in these calculations to get the maximum shifts. 

For $\vec{v}_{a}$ we consider the four flows predicted to be inside the cross-section of the fifth caustic ring\cite{CHAKRABARTY2021100838}. Velocities for the caustic ring flows are taken from Chakrabarty et al. (2021)~\cite{PhysRevD.78.063508}. The axion velocities for all four scenarios are summarized in Tab.~\ref{tab:axion_vel}. 

Inputting these values, along with the values of $\vec{v}_{D}$ detailed above, we calculate the maximum Doppler shift one could have observed over the course of Run 1B, with a $\Delta_{t}$ of 62 days between measurements of the same frequency. We opt to use this value of 62 days because it is twice the largest difference between two measurements of the same frequency. This enables a more conservative approach to estimate the shift to ensure real axion signals are not cut by using this threshold. The Doppler shifts calculated for the four flows are listed in Tab.~\ref{tab:doppler} in the main text.

\begin{center}
\begin{table}[H]
\centering
\renewcommand{\arraystretch}{1.5} 
\begin{tabular}{ |m{2.75cm}| m{1.75cm} | m{1.75cm}| m{1.75cm} | } 
\toprule[0.1ex]
\hline
\toprule[0.1ex]
\bf{Name of Flow} & $\vec{v_{a,r}}$ \bf{(km/s)} & $\vec{v}_{a,\phi}$ \bf{(km/s)} & $\vec{v_{a,z}}$ \bf{(km/s)} \\
\hline
Big  & -104 & 509 & 6.1 \\
\hline
Little  & -0.2 & 520 & 4.5 \\
\hline
Up  & -115.3 & 505.1 & 44.8 \\
\hline
Down  & -116.4 & 505.4 & -38.1 \\
\hline
\bottomrule[0.1ex]
\hline
\end{tabular}
\caption{Axion velocities used to calculate Doppler shift. Taken from \cite{CHAKRABARTY2021100838} .}
\label{tab:axion_vel}
\end{table}
\end{center}

\bibliographystyle{apsrev4-1}
\raggedright
\bibliography{references}

\begin{thebibliography}{42}%
\makeatletter
\providecommand \@ifxundefined [1]{%
 \@ifx{#1\undefined}
}%
\providecommand \@ifnum [1]{%
 \ifnum #1\expandafter \@firstoftwo
 \else \expandafter \@secondoftwo
 \fi
}%
\providecommand \@ifx [1]{%
 \ifx #1\expandafter \@firstoftwo
 \else \expandafter \@secondoftwo
 \fi
}%
\providecommand \natexlab [1]{#1}%
\providecommand \enquote  [1]{``#1''}%
\providecommand \bibnamefont  [1]{#1}%
\providecommand \bibfnamefont [1]{#1}%
\providecommand \citenamefont [1]{#1}%
\providecommand \href@noop [0]{\@secondoftwo}%
\providecommand \href [0]{\begingroup \@sanitize@url \@href}%
\providecommand \@href[1]{\@@startlink{#1}\@@href}%
\providecommand \@@href[1]{\endgroup#1\@@endlink}%
\providecommand \@sanitize@url [0]{\catcode `\\12\catcode `\$12\catcode `\&12\catcode `\#12\catcode `\^12\catcode `\_12\catcode `\%12\relax}%
\providecommand \@@startlink[1]{}%
\providecommand \@@endlink[0]{}%
\providecommand \url  [0]{\begingroup\@sanitize@url \@url }%
\providecommand \@url [1]{\endgroup\@href {#1}{\urlprefix }}%
\providecommand \urlprefix  [0]{URL }%
\providecommand \Eprint [0]{\href }%
\providecommand \doibase [0]{http://dx.doi.org/}%
\providecommand \selectlanguage [0]{\@gobble}%
\providecommand \bibinfo  [0]{\@secondoftwo}%
\providecommand \bibfield  [0]{\@secondoftwo}%
\providecommand \translation [1]{[#1]}%
\providecommand \BibitemOpen [0]{}%
\providecommand \bibitemStop [0]{}%
\providecommand \bibitemNoStop [0]{.\EOS\space}%
\providecommand \EOS [0]{\spacefactor3000\relax}%
\providecommand \BibitemShut  [1]{\csname bibitem#1\endcsname}%
\let\auto@bib@innerbib\@empty
\bibitem [{\citenamefont {Peccei}\ and\ \citenamefont {Quinn}(1977)}]{Peccei1977Sept}%
  \BibitemOpen
  \bibfield  {author} {\bibinfo {author} {\bibfnamefont {R.~D.}\ \bibnamefont {Peccei}}\ and\ \bibinfo {author} {\bibfnamefont {H.~R.}\ \bibnamefont {Quinn}},\ }\href {\doibase 10.1103/PhysRevD.16.1791} {\bibfield  {journal} {\bibinfo  {journal} {Phys. Rev. D}\ }\textbf {\bibinfo {volume} {16}},\ \bibinfo {pages} {1791} (\bibinfo {year} {1977})}\BibitemShut {NoStop}%
\bibitem [{\citenamefont {Weinberg}(1978)}]{weinberg}%
  \BibitemOpen
  \bibfield  {author} {\bibinfo {author} {\bibfnamefont {S.}~\bibnamefont {Weinberg}},\ }\href {\doibase 10.1103/PhysRevLett.40.223} {\bibfield  {journal} {\bibinfo  {journal} {Phys. Rev. Lett.}\ }\textbf {\bibinfo {volume} {40}},\ \bibinfo {pages} {223} (\bibinfo {year} {1978})}\BibitemShut {NoStop}%
\bibitem [{\citenamefont {Wilczek}(1978)}]{wilczek}%
  \BibitemOpen
  \bibfield  {author} {\bibinfo {author} {\bibfnamefont {F.}~\bibnamefont {Wilczek}},\ }\href {\doibase 10.1103/PhysRevLett.40.279} {\bibfield  {journal} {\bibinfo  {journal} {Phys. Rev. Lett.}\ }\textbf {\bibinfo {volume} {40}},\ \bibinfo {pages} {279} (\bibinfo {year} {1978})}\BibitemShut {NoStop}%
\bibitem [{\citenamefont {{Planck Collaboration}}\ \emph {et~al.}(2014)\citenamefont {{Planck Collaboration}}, \citenamefont {{Ade, P. A. R.}}, \citenamefont {{Aghanim, N.}}, \citenamefont {{Armitage-Caplan, C.}}, \citenamefont {{Arnaud, M.}}, \citenamefont {{Ashdown, M.}}, \citenamefont {{Atrio-Barandela, F.}}, \citenamefont {{Aumont, J.}}, \citenamefont {{Baccigalupi, C.}}, \citenamefont {{Banday, A. J.}}, \citenamefont {{Barreiro, R. B.}}, \citenamefont {{Bartlett, J. G.}}, \citenamefont {{Battaner, E.}}, \citenamefont {{Benabed, K.}}, \citenamefont {{Beno\^{\i}t, A.}}, \citenamefont {{Benoit-L\'evy, A.}}, \citenamefont {{Bernard, J.-P.}}, \citenamefont {{Bersanelli, M.}}, \citenamefont {{Bielewicz, P.}}, \citenamefont {{Bobin, J.}}, \citenamefont {{Bock, J. J.}}, \citenamefont {{Bonaldi, A.}}, \citenamefont {{Bond, J. R.}}, \citenamefont {{Borrill, J.}}, \citenamefont {{Bouchet, F. R.}}, \citenamefont {{Bridges, M.}}, \citenamefont {{Bucher, M.}}, \citenamefont {{Burigana, C.}}, \citenamefont {{Butler, R.
  C.}}, \citenamefont {{Calabrese, E.}}, \citenamefont {{Cappellini, B.}}, \citenamefont {{Cardoso, J.-F.}}, \citenamefont {{Catalano, A.}}, \citenamefont {{Challinor, A.}}, \citenamefont {{Chamballu, A.}}, \citenamefont {{Chary, R.-R.}}, \citenamefont {{Chen, X.}}, \citenamefont {{Chiang, H. C.}}, \citenamefont {{Chiang, L.-Y}}, \citenamefont {{Christensen, P. R.}}, \citenamefont {{Church, S.}}, \citenamefont {{Clements, D. L.}}, \citenamefont {{Colombi, S.}}, \citenamefont {{Colombo, L. P. L.}}, \citenamefont {{Couchot, F.}}, \citenamefont {{Coulais, A.}}, \citenamefont {{Crill, B. P.}}, \citenamefont {{Curto, A.}}, \citenamefont {{Cuttaia, F.}}, \citenamefont {{Danese, L.}}, \citenamefont {{Davies, R. D.}}, \citenamefont {{Davis, R. J.}}, \citenamefont {{de Bernardis, P.}}, \citenamefont {{de Rosa, A.}}, \citenamefont {{de Zotti, G.}}, \citenamefont {{Delabrouille, J.}}, \citenamefont {{Delouis, J.-M.}}, \citenamefont {{D\'esert, F.-X.}}, \citenamefont {{Dickinson, C.}}, \citenamefont {{Diego, J. M.}},
  \citenamefont {{Dolag, K.}}, \citenamefont {{Dole, H.}}, \citenamefont {{Donzelli, S.}}, \citenamefont {{Dor\'e, O.}}, \citenamefont {{Douspis, M.}}, \citenamefont {{Dunkley, J.}}, \citenamefont {{Dupac, X.}}, \citenamefont {{Efstathiou, G.}}, \citenamefont {{Elsner, F.}}, \citenamefont {{En\ss{}lin, T. A.}}, \citenamefont {{Eriksen, H. K.}}, \citenamefont {{Finelli, F.}}, \citenamefont {{Forni, O.}}, \citenamefont {{Frailis, M.}}, \citenamefont {{Fraisse, A. A.}}, \citenamefont {{Franceschi, E.}}, \citenamefont {{Gaier, T. C.}}, \citenamefont {{Galeotta, S.}}, \citenamefont {{Galli, S.}}, \citenamefont {{Ganga, K.}}, \citenamefont {{Giard, M.}}, \citenamefont {{Giardino, G.}}, \citenamefont {{Giraud-H\'eraud, Y.}}, \citenamefont {{Gjerl\o{}w, E.}}, \citenamefont {{Gonz\'alez-Nuevo, J.}}, \citenamefont {{G\'orski, K. M.}}, \citenamefont {{Gratton, S.}}, \citenamefont {{Gregorio, A.}}, \citenamefont {{Gruppuso, A.}}, \citenamefont {{Gudmundsson, J. E.}}, \citenamefont {{Haissinski, J.}}, \citenamefont
  {{Hamann, J.}}, \citenamefont {{Hansen, F. K.}}, \citenamefont {{Hanson, D.}}, \citenamefont {{Harrison, D.}}, \citenamefont {{Henrot-Versill\'e, S.}}, \citenamefont {{Hern\'andez-Monteagudo, C.}}, \citenamefont {{Herranz, D.}}, \citenamefont {{Hildebrandt, S. R.}}, \citenamefont {{Hivon, E.}}, \citenamefont {{Hobson, M.}}, \citenamefont {{Holmes, W. A.}}, \citenamefont {{Hornstrup, A.}}, \citenamefont {{Hou, Z.}}, \citenamefont {{Hovest, W.}}, \citenamefont {{Huffenberger, K. M.}}, \citenamefont {{Jaffe, A. H.}}, \citenamefont {{Jaffe, T. R.}}, \citenamefont {{Jewell, J.}}, \citenamefont {{Jones, W. C.}}, \citenamefont {{Juvela, M.}}, \citenamefont {{Keih\"anen, E.}}, \citenamefont {{Keskitalo, R.}}, \citenamefont {{Kisner, T. S.}}, \citenamefont {{Kneissl, R.}}, \citenamefont {{Knoche, J.}}, \citenamefont {{Knox, L.}}, \citenamefont {{Kunz, M.}}, \citenamefont {{Kurki-Suonio, H.}}, \citenamefont {{Lagache, G.}}, \citenamefont {{L\"ahteenm\"aki, A.}}, \citenamefont {{Lamarre, J.-M.}}, \citenamefont
  {{Lasenby, A.}}, \citenamefont {{Lattanzi, M.}}, \citenamefont {{Laureijs, R. J.}}, \citenamefont {{Lawrence, C. R.}}, \citenamefont {{Leach, S.}}, \citenamefont {{Leahy, J. P.}}, \citenamefont {{Leonardi, R.}}, \citenamefont {{Le\'on-Tavares, J.}}, \citenamefont {{Lesgourgues, J.}}, \citenamefont {{Lewis, A.}}, \citenamefont {{Liguori, M.}}, \citenamefont {{Lilje, P. B.}}, \citenamefont {{Linden-V\o{}rnle, M.}}, \citenamefont {{L\'opez-Caniego, M.}}, \citenamefont {{Lubin, P. M.}}, \citenamefont {{Mac\'{\i}as-P\'erez, J. F.}}, \citenamefont {{Maffei, B.}}, \citenamefont {{Maino, D.}}, \citenamefont {{Mandolesi, N.}}, \citenamefont {{Maris, M.}}, \citenamefont {{Marshall, D. J.}}, \citenamefont {{Martin, P. G.}}, \citenamefont {{Mart\'{\i}nez-Gonz\'alez, E.}}, \citenamefont {{Masi, S.}}, \citenamefont {{Massardi, M.}}, \citenamefont {{Matarrese, S.}}, \citenamefont {{Matthai, F.}}, \citenamefont {{Mazzotta, P.}}, \citenamefont {{Meinhold, P. R.}}, \citenamefont {{Melchiorri, A.}}, \citenamefont {{Melin,
  J.-B.}}, \citenamefont {{Mendes, L.}}, \citenamefont {{Menegoni, E.}}, \citenamefont {{Mennella, A.}}, \citenamefont {{Migliaccio, M.}}, \citenamefont {{Millea, M.}}, \citenamefont {{Mitra, S.}}, \citenamefont {{Miville-Desch\^enes, M.-A.}}, \citenamefont {{Moneti, A.}}, \citenamefont {{Montier, L.}}, \citenamefont {{Morgante, G.}}, \citenamefont {{Mortlock, D.}}, \citenamefont {{Moss, A.}}, \citenamefont {{Munshi, D.}}, \citenamefont {{Murphy, J. A.}}, \citenamefont {{Naselsky, P.}}, \citenamefont {{Nati, F.}}, \citenamefont {{Natoli, P.}}, \citenamefont {{Netterfield, C. B.}}, \citenamefont {{N\o{}rgaard-Nielsen, H. U.}}, \citenamefont {{Noviello, F.}}, \citenamefont {{Novikov, D.}}, \citenamefont {{Novikov, I.}}, \citenamefont {{O\'{}Dwyer, I. J.}}, \citenamefont {{Osborne, S.}}, \citenamefont {{Oxborrow, C. A.}}, \citenamefont {{Paci, F.}}, \citenamefont {{Pagano, L.}}, \citenamefont {{Pajot, F.}}, \citenamefont {{Paladini, R.}}, \citenamefont {{Paoletti, D.}}, \citenamefont {{Partridge, B.}},
  \citenamefont {{Pasian, F.}}, \citenamefont {{Patanchon, G.}}, \citenamefont {{Pearson, D.}}, \citenamefont {{Pearson, T. J.}}, \citenamefont {{Peiris, H. V.}}, \citenamefont {{Perdereau, O.}}, \citenamefont {{Perotto, L.}}, \citenamefont {{Perrotta, F.}}, \citenamefont {{Pettorino, V.}}, \citenamefont {{Piacentini, F.}}, \citenamefont {{Piat, M.}}, \citenamefont {{Pierpaoli, E.}}, \citenamefont {{Pietrobon, D.}}, \citenamefont {{Plaszczynski, S.}}, \citenamefont {{Platania, P.}}, \citenamefont {{Pointecouteau, E.}}, \citenamefont {{Polenta, G.}}, \citenamefont {{Ponthieu, N.}}, \citenamefont {{Popa, L.}}, \citenamefont {{Poutanen, T.}}, \citenamefont {{Pratt, G. W.}}, \citenamefont {{Pr\'ezeau, G.}}, \citenamefont {{Prunet, S.}}, \citenamefont {{Puget, J.-L.}}, \citenamefont {{Rachen, J. P.}}, \citenamefont {{Reach, W. T.}}, \citenamefont {{Rebolo, R.}}, \citenamefont {{Reinecke, M.}}, \citenamefont {{Remazeilles, M.}}, \citenamefont {{Renault, C.}}, \citenamefont {{Ricciardi, S.}}, \citenamefont {{Riller,
  T.}}, \citenamefont {{Ristorcelli, I.}}, \citenamefont {{Rocha, G.}}, \citenamefont {{Rosset, C.}}, \citenamefont {{Roudier, G.}}, \citenamefont {{Rowan-Robinson, M.}}, \citenamefont {{Rubi\~no-Mart\'{\i}n, J. A.}}, \citenamefont {{Rusholme, B.}}, \citenamefont {{Sandri, M.}}, \citenamefont {{Santos, D.}}, \citenamefont {{Savelainen, M.}}, \citenamefont {{Savini, G.}}, \citenamefont {{Scott, D.}}, \citenamefont {{Seiffert, M. D.}}, \citenamefont {{Shellard, E. P. S.}}, \citenamefont {{Spencer, L. D.}}, \citenamefont {{Starck, J.-L.}}, \citenamefont {{Stolyarov, V.}}, \citenamefont {{Stompor, R.}}, \citenamefont {{Sudiwala, R.}}, \citenamefont {{Sunyaev, R.}}, \citenamefont {{Sureau, F.}}, \citenamefont {{Sutton, D.}}, \citenamefont {{Suur-Uski, A.-S.}}, \citenamefont {{Sygnet, J.-F.}}, \citenamefont {{Tauber, J. A.}}, \citenamefont {{Tavagnacco, D.}}, \citenamefont {{Terenzi, L.}}, \citenamefont {{Toffolatti, L.}}, \citenamefont {{Tomasi, M.}}, \citenamefont {{Tristram, M.}}, \citenamefont {{Tucci, M.}},
  \citenamefont {{Tuovinen, J.}}, \citenamefont {{T\"urler, M.}}, \citenamefont {{Umana, G.}}, \citenamefont {{Valenziano, L.}}, \citenamefont {{Valiviita, J.}}, \citenamefont {{Van Tent, B.}}, \citenamefont {{Vielva, P.}}, \citenamefont {{Villa, F.}}, \citenamefont {{Vittorio, N.}}, \citenamefont {{Wade, L. A.}}, \citenamefont {{Wandelt, B. D.}}, \citenamefont {{Wehus, I. K.}}, \citenamefont {{White, M.}}, \citenamefont {{White, S. D. M.}}, \citenamefont {{Wilkinson, A.}}, \citenamefont {{Yvon, D.}}, \citenamefont {{Zacchei, A.}},\ and\ \citenamefont {{Zonca, A.}}}]{Planck}%
  \BibitemOpen
  \bibfield  {author} {\bibinfo {author} {\bibnamefont {{Planck Collaboration}}}, \bibinfo {author} {\bibnamefont {{Ade, P. A. R.}}}, \bibinfo {author} {\bibnamefont {{Aghanim, N.}}}, \bibinfo {author} {\bibnamefont {{Armitage-Caplan, C.}}}, \bibinfo {author} {\bibnamefont {{Arnaud, M.}}}, \bibinfo {author} {\bibnamefont {{Ashdown, M.}}}, \bibinfo {author} {\bibnamefont {{Atrio-Barandela, F.}}}, \bibinfo {author} {\bibnamefont {{Aumont, J.}}}, \bibinfo {author} {\bibnamefont {{Baccigalupi, C.}}}, \bibinfo {author} {\bibnamefont {{Banday, A. J.}}}, \bibinfo {author} {\bibnamefont {{Barreiro, R. B.}}}, \bibinfo {author} {\bibnamefont {{Bartlett, J. G.}}}, \bibinfo {author} {\bibnamefont {{Battaner, E.}}}, \bibinfo {author} {\bibnamefont {{Benabed, K.}}}, \bibinfo {author} {\bibnamefont {{Beno\^{\i}t, A.}}}, \bibinfo {author} {\bibnamefont {{Benoit-L\'evy, A.}}}, \bibinfo {author} {\bibnamefont {{Bernard, J.-P.}}}, \bibinfo {author} {\bibnamefont {{Bersanelli, M.}}}, \bibinfo {author} {\bibnamefont {{Bielewicz,
  P.}}}, \bibinfo {author} {\bibnamefont {{Bobin, J.}}}, \bibinfo {author} {\bibnamefont {{Bock, J. J.}}}, \bibinfo {author} {\bibnamefont {{Bonaldi, A.}}}, \bibinfo {author} {\bibnamefont {{Bond, J. R.}}}, \bibinfo {author} {\bibnamefont {{Borrill, J.}}}, \bibinfo {author} {\bibnamefont {{Bouchet, F. R.}}}, \bibinfo {author} {\bibnamefont {{Bridges, M.}}}, \bibinfo {author} {\bibnamefont {{Bucher, M.}}}, \bibinfo {author} {\bibnamefont {{Burigana, C.}}}, \bibinfo {author} {\bibnamefont {{Butler, R. C.}}}, \bibinfo {author} {\bibnamefont {{Calabrese, E.}}}, \bibinfo {author} {\bibnamefont {{Cappellini, B.}}}, \bibinfo {author} {\bibnamefont {{Cardoso, J.-F.}}}, \bibinfo {author} {\bibnamefont {{Catalano, A.}}}, \bibinfo {author} {\bibnamefont {{Challinor, A.}}}, \bibinfo {author} {\bibnamefont {{Chamballu, A.}}}, \bibinfo {author} {\bibnamefont {{Chary, R.-R.}}}, \bibinfo {author} {\bibnamefont {{Chen, X.}}}, \bibinfo {author} {\bibnamefont {{Chiang, H. C.}}}, \bibinfo {author} {\bibnamefont {{Chiang,
  L.-Y}}}, \bibinfo {author} {\bibnamefont {{Christensen, P. R.}}}, \bibinfo {author} {\bibnamefont {{Church, S.}}}, \bibinfo {author} {\bibnamefont {{Clements, D. L.}}}, \bibinfo {author} {\bibnamefont {{Colombi, S.}}}, \bibinfo {author} {\bibnamefont {{Colombo, L. P. L.}}}, \bibinfo {author} {\bibnamefont {{Couchot, F.}}}, \bibinfo {author} {\bibnamefont {{Coulais, A.}}}, \bibinfo {author} {\bibnamefont {{Crill, B. P.}}}, \bibinfo {author} {\bibnamefont {{Curto, A.}}}, \bibinfo {author} {\bibnamefont {{Cuttaia, F.}}}, \bibinfo {author} {\bibnamefont {{Danese, L.}}}, \bibinfo {author} {\bibnamefont {{Davies, R. D.}}}, \bibinfo {author} {\bibnamefont {{Davis, R. J.}}}, \bibinfo {author} {\bibnamefont {{de Bernardis, P.}}}, \bibinfo {author} {\bibnamefont {{de Rosa, A.}}}, \bibinfo {author} {\bibnamefont {{de Zotti, G.}}}, \bibinfo {author} {\bibnamefont {{Delabrouille, J.}}}, \bibinfo {author} {\bibnamefont {{Delouis, J.-M.}}}, \bibinfo {author} {\bibnamefont {{D\'esert, F.-X.}}}, \bibinfo {author}
  {\bibnamefont {{Dickinson, C.}}}, \bibinfo {author} {\bibnamefont {{Diego, J. M.}}}, \bibinfo {author} {\bibnamefont {{Dolag, K.}}}, \bibinfo {author} {\bibnamefont {{Dole, H.}}}, \bibinfo {author} {\bibnamefont {{Donzelli, S.}}}, \bibinfo {author} {\bibnamefont {{Dor\'e, O.}}}, \bibinfo {author} {\bibnamefont {{Douspis, M.}}}, \bibinfo {author} {\bibnamefont {{Dunkley, J.}}}, \bibinfo {author} {\bibnamefont {{Dupac, X.}}}, \bibinfo {author} {\bibnamefont {{Efstathiou, G.}}}, \bibinfo {author} {\bibnamefont {{Elsner, F.}}}, \bibinfo {author} {\bibnamefont {{En\ss{}lin, T. A.}}}, \bibinfo {author} {\bibnamefont {{Eriksen, H. K.}}}, \bibinfo {author} {\bibnamefont {{Finelli, F.}}}, \bibinfo {author} {\bibnamefont {{Forni, O.}}}, \bibinfo {author} {\bibnamefont {{Frailis, M.}}}, \bibinfo {author} {\bibnamefont {{Fraisse, A. A.}}}, \bibinfo {author} {\bibnamefont {{Franceschi, E.}}}, \bibinfo {author} {\bibnamefont {{Gaier, T. C.}}}, \bibinfo {author} {\bibnamefont {{Galeotta, S.}}}, \bibinfo {author}
  {\bibnamefont {{Galli, S.}}}, \bibinfo {author} {\bibnamefont {{Ganga, K.}}}, \bibinfo {author} {\bibnamefont {{Giard, M.}}}, \bibinfo {author} {\bibnamefont {{Giardino, G.}}}, \bibinfo {author} {\bibnamefont {{Giraud-H\'eraud, Y.}}}, \bibinfo {author} {\bibnamefont {{Gjerl\o{}w, E.}}}, \bibinfo {author} {\bibnamefont {{Gonz\'alez-Nuevo, J.}}}, \bibinfo {author} {\bibnamefont {{G\'orski, K. M.}}}, \bibinfo {author} {\bibnamefont {{Gratton, S.}}}, \bibinfo {author} {\bibnamefont {{Gregorio, A.}}}, \bibinfo {author} {\bibnamefont {{Gruppuso, A.}}}, \bibinfo {author} {\bibnamefont {{Gudmundsson, J. E.}}}, \bibinfo {author} {\bibnamefont {{Haissinski, J.}}}, \bibinfo {author} {\bibnamefont {{Hamann, J.}}}, \bibinfo {author} {\bibnamefont {{Hansen, F. K.}}}, \bibinfo {author} {\bibnamefont {{Hanson, D.}}}, \bibinfo {author} {\bibnamefont {{Harrison, D.}}}, \bibinfo {author} {\bibnamefont {{Henrot-Versill\'e, S.}}}, \bibinfo {author} {\bibnamefont {{Hern\'andez-Monteagudo, C.}}}, \bibinfo {author} {\bibnamefont
  {{Herranz, D.}}}, \bibinfo {author} {\bibnamefont {{Hildebrandt, S. R.}}}, \bibinfo {author} {\bibnamefont {{Hivon, E.}}}, \bibinfo {author} {\bibnamefont {{Hobson, M.}}}, \bibinfo {author} {\bibnamefont {{Holmes, W. A.}}}, \bibinfo {author} {\bibnamefont {{Hornstrup, A.}}}, \bibinfo {author} {\bibnamefont {{Hou, Z.}}}, \bibinfo {author} {\bibnamefont {{Hovest, W.}}}, \bibinfo {author} {\bibnamefont {{Huffenberger, K. M.}}}, \bibinfo {author} {\bibnamefont {{Jaffe, A. H.}}}, \bibinfo {author} {\bibnamefont {{Jaffe, T. R.}}}, \bibinfo {author} {\bibnamefont {{Jewell, J.}}}, \bibinfo {author} {\bibnamefont {{Jones, W. C.}}}, \bibinfo {author} {\bibnamefont {{Juvela, M.}}}, \bibinfo {author} {\bibnamefont {{Keih\"anen, E.}}}, \bibinfo {author} {\bibnamefont {{Keskitalo, R.}}}, \bibinfo {author} {\bibnamefont {{Kisner, T. S.}}}, \bibinfo {author} {\bibnamefont {{Kneissl, R.}}}, \bibinfo {author} {\bibnamefont {{Knoche, J.}}}, \bibinfo {author} {\bibnamefont {{Knox, L.}}}, \bibinfo {author} {\bibnamefont {{Kunz,
  M.}}}, \bibinfo {author} {\bibnamefont {{Kurki-Suonio, H.}}}, \bibinfo {author} {\bibnamefont {{Lagache, G.}}}, \bibinfo {author} {\bibnamefont {{L\"ahteenm\"aki, A.}}}, \bibinfo {author} {\bibnamefont {{Lamarre, J.-M.}}}, \bibinfo {author} {\bibnamefont {{Lasenby, A.}}}, \bibinfo {author} {\bibnamefont {{Lattanzi, M.}}}, \bibinfo {author} {\bibnamefont {{Laureijs, R. J.}}}, \bibinfo {author} {\bibnamefont {{Lawrence, C. R.}}}, \bibinfo {author} {\bibnamefont {{Leach, S.}}}, \bibinfo {author} {\bibnamefont {{Leahy, J. P.}}}, \bibinfo {author} {\bibnamefont {{Leonardi, R.}}}, \bibinfo {author} {\bibnamefont {{Le\'on-Tavares, J.}}}, \bibinfo {author} {\bibnamefont {{Lesgourgues, J.}}}, \bibinfo {author} {\bibnamefont {{Lewis, A.}}}, \bibinfo {author} {\bibnamefont {{Liguori, M.}}}, \bibinfo {author} {\bibnamefont {{Lilje, P. B.}}}, \bibinfo {author} {\bibnamefont {{Linden-V\o{}rnle, M.}}}, \bibinfo {author} {\bibnamefont {{L\'opez-Caniego, M.}}}, \bibinfo {author} {\bibnamefont {{Lubin, P. M.}}}, \bibinfo
  {author} {\bibnamefont {{Mac\'{\i}as-P\'erez, J. F.}}}, \bibinfo {author} {\bibnamefont {{Maffei, B.}}}, \bibinfo {author} {\bibnamefont {{Maino, D.}}}, \bibinfo {author} {\bibnamefont {{Mandolesi, N.}}}, \bibinfo {author} {\bibnamefont {{Maris, M.}}}, \bibinfo {author} {\bibnamefont {{Marshall, D. J.}}}, \bibinfo {author} {\bibnamefont {{Martin, P. G.}}}, \bibinfo {author} {\bibnamefont {{Mart\'{\i}nez-Gonz\'alez, E.}}}, \bibinfo {author} {\bibnamefont {{Masi, S.}}}, \bibinfo {author} {\bibnamefont {{Massardi, M.}}}, \bibinfo {author} {\bibnamefont {{Matarrese, S.}}}, \bibinfo {author} {\bibnamefont {{Matthai, F.}}}, \bibinfo {author} {\bibnamefont {{Mazzotta, P.}}}, \bibinfo {author} {\bibnamefont {{Meinhold, P. R.}}}, \bibinfo {author} {\bibnamefont {{Melchiorri, A.}}}, \bibinfo {author} {\bibnamefont {{Melin, J.-B.}}}, \bibinfo {author} {\bibnamefont {{Mendes, L.}}}, \bibinfo {author} {\bibnamefont {{Menegoni, E.}}}, \bibinfo {author} {\bibnamefont {{Mennella, A.}}}, \bibinfo {author} {\bibnamefont
  {{Migliaccio, M.}}}, \bibinfo {author} {\bibnamefont {{Millea, M.}}}, \bibinfo {author} {\bibnamefont {{Mitra, S.}}}, \bibinfo {author} {\bibnamefont {{Miville-Desch\^enes, M.-A.}}}, \bibinfo {author} {\bibnamefont {{Moneti, A.}}}, \bibinfo {author} {\bibnamefont {{Montier, L.}}}, \bibinfo {author} {\bibnamefont {{Morgante, G.}}}, \bibinfo {author} {\bibnamefont {{Mortlock, D.}}}, \bibinfo {author} {\bibnamefont {{Moss, A.}}}, \bibinfo {author} {\bibnamefont {{Munshi, D.}}}, \bibinfo {author} {\bibnamefont {{Murphy, J. A.}}}, \bibinfo {author} {\bibnamefont {{Naselsky, P.}}}, \bibinfo {author} {\bibnamefont {{Nati, F.}}}, \bibinfo {author} {\bibnamefont {{Natoli, P.}}}, \bibinfo {author} {\bibnamefont {{Netterfield, C. B.}}}, \bibinfo {author} {\bibnamefont {{N\o{}rgaard-Nielsen, H. U.}}}, \bibinfo {author} {\bibnamefont {{Noviello, F.}}}, \bibinfo {author} {\bibnamefont {{Novikov, D.}}}, \bibinfo {author} {\bibnamefont {{Novikov, I.}}}, \bibinfo {author} {\bibnamefont {{O\'{}Dwyer, I. J.}}}, \bibinfo
  {author} {\bibnamefont {{Osborne, S.}}}, \bibinfo {author} {\bibnamefont {{Oxborrow, C. A.}}}, \bibinfo {author} {\bibnamefont {{Paci, F.}}}, \bibinfo {author} {\bibnamefont {{Pagano, L.}}}, \bibinfo {author} {\bibnamefont {{Pajot, F.}}}, \bibinfo {author} {\bibnamefont {{Paladini, R.}}}, \bibinfo {author} {\bibnamefont {{Paoletti, D.}}}, \bibinfo {author} {\bibnamefont {{Partridge, B.}}}, \bibinfo {author} {\bibnamefont {{Pasian, F.}}}, \bibinfo {author} {\bibnamefont {{Patanchon, G.}}}, \bibinfo {author} {\bibnamefont {{Pearson, D.}}}, \bibinfo {author} {\bibnamefont {{Pearson, T. J.}}}, \bibinfo {author} {\bibnamefont {{Peiris, H. V.}}}, \bibinfo {author} {\bibnamefont {{Perdereau, O.}}}, \bibinfo {author} {\bibnamefont {{Perotto, L.}}}, \bibinfo {author} {\bibnamefont {{Perrotta, F.}}}, \bibinfo {author} {\bibnamefont {{Pettorino, V.}}}, \bibinfo {author} {\bibnamefont {{Piacentini, F.}}}, \bibinfo {author} {\bibnamefont {{Piat, M.}}}, \bibinfo {author} {\bibnamefont {{Pierpaoli, E.}}}, \bibinfo
  {author} {\bibnamefont {{Pietrobon, D.}}}, \bibinfo {author} {\bibnamefont {{Plaszczynski, S.}}}, \bibinfo {author} {\bibnamefont {{Platania, P.}}}, \bibinfo {author} {\bibnamefont {{Pointecouteau, E.}}}, \bibinfo {author} {\bibnamefont {{Polenta, G.}}}, \bibinfo {author} {\bibnamefont {{Ponthieu, N.}}}, \bibinfo {author} {\bibnamefont {{Popa, L.}}}, \bibinfo {author} {\bibnamefont {{Poutanen, T.}}}, \bibinfo {author} {\bibnamefont {{Pratt, G. W.}}}, \bibinfo {author} {\bibnamefont {{Pr\'ezeau, G.}}}, \bibinfo {author} {\bibnamefont {{Prunet, S.}}}, \bibinfo {author} {\bibnamefont {{Puget, J.-L.}}}, \bibinfo {author} {\bibnamefont {{Rachen, J. P.}}}, \bibinfo {author} {\bibnamefont {{Reach, W. T.}}}, \bibinfo {author} {\bibnamefont {{Rebolo, R.}}}, \bibinfo {author} {\bibnamefont {{Reinecke, M.}}}, \bibinfo {author} {\bibnamefont {{Remazeilles, M.}}}, \bibinfo {author} {\bibnamefont {{Renault, C.}}}, \bibinfo {author} {\bibnamefont {{Ricciardi, S.}}}, \bibinfo {author} {\bibnamefont {{Riller, T.}}},
  \bibinfo {author} {\bibnamefont {{Ristorcelli, I.}}}, \bibinfo {author} {\bibnamefont {{Rocha, G.}}}, \bibinfo {author} {\bibnamefont {{Rosset, C.}}}, \bibinfo {author} {\bibnamefont {{Roudier, G.}}}, \bibinfo {author} {\bibnamefont {{Rowan-Robinson, M.}}}, \bibinfo {author} {\bibnamefont {{Rubi\~no-Mart\'{\i}n, J. A.}}}, \bibinfo {author} {\bibnamefont {{Rusholme, B.}}}, \bibinfo {author} {\bibnamefont {{Sandri, M.}}}, \bibinfo {author} {\bibnamefont {{Santos, D.}}}, \bibinfo {author} {\bibnamefont {{Savelainen, M.}}}, \bibinfo {author} {\bibnamefont {{Savini, G.}}}, \bibinfo {author} {\bibnamefont {{Scott, D.}}}, \bibinfo {author} {\bibnamefont {{Seiffert, M. D.}}}, \bibinfo {author} {\bibnamefont {{Shellard, E. P. S.}}}, \bibinfo {author} {\bibnamefont {{Spencer, L. D.}}}, \bibinfo {author} {\bibnamefont {{Starck, J.-L.}}}, \bibinfo {author} {\bibnamefont {{Stolyarov, V.}}}, \bibinfo {author} {\bibnamefont {{Stompor, R.}}}, \bibinfo {author} {\bibnamefont {{Sudiwala, R.}}}, \bibinfo {author}
  {\bibnamefont {{Sunyaev, R.}}}, \bibinfo {author} {\bibnamefont {{Sureau, F.}}}, \bibinfo {author} {\bibnamefont {{Sutton, D.}}}, \bibinfo {author} {\bibnamefont {{Suur-Uski, A.-S.}}}, \bibinfo {author} {\bibnamefont {{Sygnet, J.-F.}}}, \bibinfo {author} {\bibnamefont {{Tauber, J. A.}}}, \bibinfo {author} {\bibnamefont {{Tavagnacco, D.}}}, \bibinfo {author} {\bibnamefont {{Terenzi, L.}}}, \bibinfo {author} {\bibnamefont {{Toffolatti, L.}}}, \bibinfo {author} {\bibnamefont {{Tomasi, M.}}}, \bibinfo {author} {\bibnamefont {{Tristram, M.}}}, \bibinfo {author} {\bibnamefont {{Tucci, M.}}}, \bibinfo {author} {\bibnamefont {{Tuovinen, J.}}}, \bibinfo {author} {\bibnamefont {{T\"urler, M.}}}, \bibinfo {author} {\bibnamefont {{Umana, G.}}}, \bibinfo {author} {\bibnamefont {{Valenziano, L.}}}, \bibinfo {author} {\bibnamefont {{Valiviita, J.}}}, \bibinfo {author} {\bibnamefont {{Van Tent, B.}}}, \bibinfo {author} {\bibnamefont {{Vielva, P.}}}, \bibinfo {author} {\bibnamefont {{Villa, F.}}}, \bibinfo {author}
  {\bibnamefont {{Vittorio, N.}}}, \bibinfo {author} {\bibnamefont {{Wade, L. A.}}}, \bibinfo {author} {\bibnamefont {{Wandelt, B. D.}}}, \bibinfo {author} {\bibnamefont {{Wehus, I. K.}}}, \bibinfo {author} {\bibnamefont {{White, M.}}}, \bibinfo {author} {\bibnamefont {{White, S. D. M.}}}, \bibinfo {author} {\bibnamefont {{Wilkinson, A.}}}, \bibinfo {author} {\bibnamefont {{Yvon, D.}}}, \bibinfo {author} {\bibnamefont {{Zacchei, A.}}}, \ and\ \bibinfo {author} {\bibnamefont {{Zonca, A.}}},\ }\href {\doibase 10.1051/0004-6361/201321591} {\bibfield  {journal} {\bibinfo  {journal} {A\&A}\ }\textbf {\bibinfo {volume} {571}},\ \bibinfo {pages} {A16} (\bibinfo {year} {2014})}\BibitemShut {NoStop}%
\bibitem [{\citenamefont {Abbott}\ and\ \citenamefont {Sikivie}(1983)}]{ABBOTT1983133}%
  \BibitemOpen
  \bibfield  {author} {\bibinfo {author} {\bibfnamefont {L.}~\bibnamefont {Abbott}}\ and\ \bibinfo {author} {\bibfnamefont {P.}~\bibnamefont {Sikivie}},\ }\href {\doibase https://doi.org/10.1016/0370-2693(83)90638-X} {\bibfield  {journal} {\bibinfo  {journal} {Physics Letters B}\ }\textbf {\bibinfo {volume} {120}},\ \bibinfo {pages} {133} (\bibinfo {year} {1983})}\BibitemShut {NoStop}%
\bibitem [{\citenamefont {Dine}\ and\ \citenamefont {Fischler}(1983)}]{DINE1983}%
  \BibitemOpen
  \bibfield  {author} {\bibinfo {author} {\bibfnamefont {M.}~\bibnamefont {Dine}}\ and\ \bibinfo {author} {\bibfnamefont {W.}~\bibnamefont {Fischler}},\ }\href {\doibase https://doi.org/10.1016/0370-2693(83)90639-1} {\bibfield  {journal} {\bibinfo  {journal} {Physics Letters B}\ }\textbf {\bibinfo {volume} {120}},\ \bibinfo {pages} {137} (\bibinfo {year} {1983})}\BibitemShut {NoStop}%
\bibitem [{\citenamefont {Preskill}\ \emph {et~al.}(1983)\citenamefont {Preskill}, \citenamefont {Wise},\ and\ \citenamefont {Wilczek}}]{PRESKILL1983}%
  \BibitemOpen
  \bibfield  {author} {\bibinfo {author} {\bibfnamefont {J.}~\bibnamefont {Preskill}}, \bibinfo {author} {\bibfnamefont {M.~B.}\ \bibnamefont {Wise}}, \ and\ \bibinfo {author} {\bibfnamefont {F.}~\bibnamefont {Wilczek}},\ }\href {\doibase https://doi.org/10.1016/0370-2693(83)90637-8} {\bibfield  {journal} {\bibinfo  {journal} {Physics Letters B}\ }\textbf {\bibinfo {volume} {120}},\ \bibinfo {pages} {127} (\bibinfo {year} {1983})}\BibitemShut {NoStop}%
\bibitem [{\citenamefont {Ipser}\ and\ \citenamefont {Sikivie}(1983)}]{Ipser1983}%
  \BibitemOpen
  \bibfield  {author} {\bibinfo {author} {\bibfnamefont {J.}~\bibnamefont {Ipser}}\ and\ \bibinfo {author} {\bibfnamefont {P.}~\bibnamefont {Sikivie}},\ }\href {\doibase 10.1103/PhysRevLett.50.925} {\bibfield  {journal} {\bibinfo  {journal} {Phys. Rev. Lett.}\ }\textbf {\bibinfo {volume} {50}},\ \bibinfo {pages} {925} (\bibinfo {year} {1983})}\BibitemShut {NoStop}%
\bibitem [{\citenamefont {Sikivie}(1983)}]{sikiviehaloscope}%
  \BibitemOpen
  \bibfield  {author} {\bibinfo {author} {\bibfnamefont {P.}~\bibnamefont {Sikivie}},\ }\href {\doibase 10.1103/PhysRevLett.51.1415} {\bibfield  {journal} {\bibinfo  {journal} {Phys. Rev. Lett.}\ }\textbf {\bibinfo {volume} {51}},\ \bibinfo {pages} {1415} (\bibinfo {year} {1983})}\BibitemShut {NoStop}%
\bibitem [{\citenamefont {Kim}(1979)}]{KIM1979}%
  \BibitemOpen
  \bibfield  {author} {\bibinfo {author} {\bibfnamefont {J.~E.}\ \bibnamefont {Kim}},\ }\href {\doibase 10.1103/PhysRevLett.43.103} {\bibfield  {journal} {\bibinfo  {journal} {Phys. Rev. Lett.}\ }\textbf {\bibinfo {volume} {43}},\ \bibinfo {pages} {103} (\bibinfo {year} {1979})}\BibitemShut {NoStop}%
\bibitem [{\citenamefont {Shifman}\ \emph {et~al.}(1980)\citenamefont {Shifman}, \citenamefont {Vainshtein},\ and\ \citenamefont {Zakharov}}]{SHIFMAN1980}%
  \BibitemOpen
  \bibfield  {author} {\bibinfo {author} {\bibfnamefont {M.}~\bibnamefont {Shifman}}, \bibinfo {author} {\bibfnamefont {A.}~\bibnamefont {Vainshtein}}, \ and\ \bibinfo {author} {\bibfnamefont {V.}~\bibnamefont {Zakharov}},\ }\href {\doibase https://doi.org/10.1016/0550-3213(80)90209-6} {\bibfield  {journal} {\bibinfo  {journal} {Nuclear Physics B}\ }\textbf {\bibinfo {volume} {166}},\ \bibinfo {pages} {493} (\bibinfo {year} {1980})}\BibitemShut {NoStop}%
\bibitem [{\citenamefont {Dine}\ \emph {et~al.}(1981)\citenamefont {Dine}, \citenamefont {Fischler},\ and\ \citenamefont {Srednicki}}]{DINE1981}%
  \BibitemOpen
  \bibfield  {author} {\bibinfo {author} {\bibfnamefont {M.}~\bibnamefont {Dine}}, \bibinfo {author} {\bibfnamefont {W.}~\bibnamefont {Fischler}}, \ and\ \bibinfo {author} {\bibfnamefont {M.}~\bibnamefont {Srednicki}},\ }\href {\doibase https://doi.org/10.1016/0370-2693(81)90590-6} {\bibfield  {journal} {\bibinfo  {journal} {Physics Letters B}\ }\textbf {\bibinfo {volume} {104}},\ \bibinfo {pages} {199} (\bibinfo {year} {1981})}\BibitemShut {NoStop}%
\bibitem [{\citenamefont {Zhitnitsky}(1980)}]{Zhitnitsky:1980tq}%
  \BibitemOpen
  \bibfield  {author} {\bibinfo {author} {\bibfnamefont {A.~R.}\ \bibnamefont {Zhitnitsky}},\ }\href@noop {} {\bibfield  {journal} {\bibinfo  {journal} {Sov. J. Nucl. Phys.}\ }\textbf {\bibinfo {volume} {31}},\ \bibinfo {pages} {260} (\bibinfo {year} {1980})}\BibitemShut {NoStop}%
\bibitem [{\citenamefont {Du}\ \emph {et~al.}(2018)\citenamefont {Du}, \citenamefont {Force}, \citenamefont {Khatiwada}, \citenamefont {Lentz}, \citenamefont {Ottens}, \citenamefont {Rosenberg}, \citenamefont {Rybka}, \citenamefont {Carosi}, \citenamefont {Woollett}, \citenamefont {Bowring},\ and\ \citenamefont {et~al.}}]{Du_2018}%
  \BibitemOpen
  \bibfield  {author} {\bibinfo {author} {\bibfnamefont {N.}~\bibnamefont {Du}}, \bibinfo {author} {\bibfnamefont {N.}~\bibnamefont {Force}}, \bibinfo {author} {\bibfnamefont {R.}~\bibnamefont {Khatiwada}}, \bibinfo {author} {\bibfnamefont {E.}~\bibnamefont {Lentz}}, \bibinfo {author} {\bibfnamefont {R.}~\bibnamefont {Ottens}}, \bibinfo {author} {\bibfnamefont {L.}~\bibnamefont {Rosenberg}}, \bibinfo {author} {\bibfnamefont {G.}~\bibnamefont {Rybka}}, \bibinfo {author} {\bibfnamefont {G.}~\bibnamefont {Carosi}}, \bibinfo {author} {\bibfnamefont {N.}~\bibnamefont {Woollett}}, \bibinfo {author} {\bibfnamefont {D.}~\bibnamefont {Bowring}}, \ and\ \bibinfo {author} {\bibnamefont {et~al.}},\ }\href {\doibase 10.1103/physrevlett.120.151301} {\bibfield  {journal} {\bibinfo  {journal} {Phys. Rev. Lett.}\ }\textbf {\bibinfo {volume} {120}} (\bibinfo {year} {2018}),\ 10.1103/physrevlett.120.151301}\BibitemShut {NoStop}%
\bibitem [{\citenamefont {Braine}\ \emph {et~al.}(2020)\citenamefont {Braine}, \citenamefont {Cervantes}, \citenamefont {Crisosto}, \citenamefont {Du}, \citenamefont {Kimes}, \citenamefont {Rosenberg}, \citenamefont {Rybka}, \citenamefont {Yang}, \citenamefont {Bowring}, \citenamefont {Chou}, \citenamefont {Khatiwada}, \citenamefont {Sonnenschein}, \citenamefont {Wester}, \citenamefont {Carosi}, \citenamefont {Woollett}, \citenamefont {Duffy}, \citenamefont {Bradley}, \citenamefont {Boutan}, \citenamefont {Jones}, \citenamefont {LaRoque}, \citenamefont {Oblath}, \citenamefont {Taubman}, \citenamefont {Clarke}, \citenamefont {Dove}, \citenamefont {Eddins}, \citenamefont {O'Kelley}, \citenamefont {Nawaz}, \citenamefont {Siddiqi}, \citenamefont {Stevenson}, \citenamefont {Agrawal}, \citenamefont {Dixit}, \citenamefont {Gleason}, \citenamefont {Jois}, \citenamefont {Sikivie}, \citenamefont {Solomon}, \citenamefont {Sullivan}, \citenamefont {Tanner}, \citenamefont {Lentz}, \citenamefont {Daw}, \citenamefont
  {Buckley}, \citenamefont {Harrington}, \citenamefont {Henriksen},\ and\ \citenamefont {Murch}}]{Run1B_Full}%
  \BibitemOpen
  \bibfield  {author} {\bibinfo {author} {\bibfnamefont {T.}~\bibnamefont {Braine}}, \bibinfo {author} {\bibfnamefont {R.}~\bibnamefont {Cervantes}}, \bibinfo {author} {\bibfnamefont {N.}~\bibnamefont {Crisosto}}, \bibinfo {author} {\bibfnamefont {N.}~\bibnamefont {Du}}, \bibinfo {author} {\bibfnamefont {S.}~\bibnamefont {Kimes}}, \bibinfo {author} {\bibfnamefont {L.~J.}\ \bibnamefont {Rosenberg}}, \bibinfo {author} {\bibfnamefont {G.}~\bibnamefont {Rybka}}, \bibinfo {author} {\bibfnamefont {J.}~\bibnamefont {Yang}}, \bibinfo {author} {\bibfnamefont {D.}~\bibnamefont {Bowring}}, \bibinfo {author} {\bibfnamefont {A.~S.}\ \bibnamefont {Chou}}, \bibinfo {author} {\bibfnamefont {R.}~\bibnamefont {Khatiwada}}, \bibinfo {author} {\bibfnamefont {A.}~\bibnamefont {Sonnenschein}}, \bibinfo {author} {\bibfnamefont {W.}~\bibnamefont {Wester}}, \bibinfo {author} {\bibfnamefont {G.}~\bibnamefont {Carosi}}, \bibinfo {author} {\bibfnamefont {N.}~\bibnamefont {Woollett}}, \bibinfo {author} {\bibfnamefont {L.~D.}\
  \bibnamefont {Duffy}}, \bibinfo {author} {\bibfnamefont {R.}~\bibnamefont {Bradley}}, \bibinfo {author} {\bibfnamefont {C.}~\bibnamefont {Boutan}}, \bibinfo {author} {\bibfnamefont {M.}~\bibnamefont {Jones}}, \bibinfo {author} {\bibfnamefont {B.~H.}\ \bibnamefont {LaRoque}}, \bibinfo {author} {\bibfnamefont {N.~S.}\ \bibnamefont {Oblath}}, \bibinfo {author} {\bibfnamefont {M.~S.}\ \bibnamefont {Taubman}}, \bibinfo {author} {\bibfnamefont {J.}~\bibnamefont {Clarke}}, \bibinfo {author} {\bibfnamefont {A.}~\bibnamefont {Dove}}, \bibinfo {author} {\bibfnamefont {A.}~\bibnamefont {Eddins}}, \bibinfo {author} {\bibfnamefont {S.~R.}\ \bibnamefont {O'Kelley}}, \bibinfo {author} {\bibfnamefont {S.}~\bibnamefont {Nawaz}}, \bibinfo {author} {\bibfnamefont {I.}~\bibnamefont {Siddiqi}}, \bibinfo {author} {\bibfnamefont {N.}~\bibnamefont {Stevenson}}, \bibinfo {author} {\bibfnamefont {A.}~\bibnamefont {Agrawal}}, \bibinfo {author} {\bibfnamefont {A.~V.}\ \bibnamefont {Dixit}}, \bibinfo {author} {\bibfnamefont {J.~R.}\
  \bibnamefont {Gleason}}, \bibinfo {author} {\bibfnamefont {S.}~\bibnamefont {Jois}}, \bibinfo {author} {\bibfnamefont {P.}~\bibnamefont {Sikivie}}, \bibinfo {author} {\bibfnamefont {J.~A.}\ \bibnamefont {Solomon}}, \bibinfo {author} {\bibfnamefont {N.~S.}\ \bibnamefont {Sullivan}}, \bibinfo {author} {\bibfnamefont {D.~B.}\ \bibnamefont {Tanner}}, \bibinfo {author} {\bibfnamefont {E.}~\bibnamefont {Lentz}}, \bibinfo {author} {\bibfnamefont {E.~J.}\ \bibnamefont {Daw}}, \bibinfo {author} {\bibfnamefont {J.~H.}\ \bibnamefont {Buckley}}, \bibinfo {author} {\bibfnamefont {P.~M.}\ \bibnamefont {Harrington}}, \bibinfo {author} {\bibfnamefont {E.~A.}\ \bibnamefont {Henriksen}}, \ and\ \bibinfo {author} {\bibfnamefont {K.~W.}\ \bibnamefont {Murch}} (\bibinfo {collaboration} {ADMX Collaboration}),\ }\href {\doibase 10.1103/PhysRevLett.124.101303} {\bibfield  {journal} {\bibinfo  {journal} {Phys. Rev. Lett.}\ }\textbf {\bibinfo {volume} {124}},\ \bibinfo {pages} {101303} (\bibinfo {year} {2020})}\BibitemShut {NoStop}%
\bibitem [{\citenamefont {Bartram}\ \emph {et~al.}(2021{\natexlab{a}})\citenamefont {Bartram}, \citenamefont {Braine}, \citenamefont {Burns}, \citenamefont {Cervantes}, \citenamefont {Crisosto}, \citenamefont {Du}, \citenamefont {Korandla}, \citenamefont {Leum}, \citenamefont {Mohapatra}, \citenamefont {Nitta}, \citenamefont {Rosenberg}, \citenamefont {Rybka}, \citenamefont {Yang}, \citenamefont {Clarke}, \citenamefont {Siddiqi}, \citenamefont {Agrawal}, \citenamefont {Dixit}, \citenamefont {Awida}, \citenamefont {Chou}, \citenamefont {Hollister}, \citenamefont {Knirck}, \citenamefont {Sonnenschein}, \citenamefont {Wester}, \citenamefont {Gleason}, \citenamefont {Hipp}, \citenamefont {Jois}, \citenamefont {Sikivie}, \citenamefont {Sullivan}, \citenamefont {Tanner}, \citenamefont {Lentz}, \citenamefont {Khatiwada}, \citenamefont {Carosi}, \citenamefont {Robertson}, \citenamefont {Woollett}, \citenamefont {Duffy}, \citenamefont {Boutan}, \citenamefont {Jones}, \citenamefont {LaRoque}, \citenamefont {Oblath},
  \citenamefont {Taubman}, \citenamefont {Daw}, \citenamefont {Perry}, \citenamefont {Buckley}, \citenamefont {Gaikwad}, \citenamefont {Hoffman}, \citenamefont {Murch}, \citenamefont {Goryachev}, \citenamefont {McAllister}, \citenamefont {Quiskamp}, \citenamefont {Thomson},\ and\ \citenamefont {Tobar}}]{Run1C_P1}%
  \BibitemOpen
  \bibfield  {author} {\bibinfo {author} {\bibfnamefont {C.}~\bibnamefont {Bartram}}, \bibinfo {author} {\bibfnamefont {T.}~\bibnamefont {Braine}}, \bibinfo {author} {\bibfnamefont {E.}~\bibnamefont {Burns}}, \bibinfo {author} {\bibfnamefont {R.}~\bibnamefont {Cervantes}}, \bibinfo {author} {\bibfnamefont {N.}~\bibnamefont {Crisosto}}, \bibinfo {author} {\bibfnamefont {N.}~\bibnamefont {Du}}, \bibinfo {author} {\bibfnamefont {H.}~\bibnamefont {Korandla}}, \bibinfo {author} {\bibfnamefont {G.}~\bibnamefont {Leum}}, \bibinfo {author} {\bibfnamefont {P.}~\bibnamefont {Mohapatra}}, \bibinfo {author} {\bibfnamefont {T.}~\bibnamefont {Nitta}}, \bibinfo {author} {\bibfnamefont {L.~J.}\ \bibnamefont {Rosenberg}}, \bibinfo {author} {\bibfnamefont {G.}~\bibnamefont {Rybka}}, \bibinfo {author} {\bibfnamefont {J.}~\bibnamefont {Yang}}, \bibinfo {author} {\bibfnamefont {J.}~\bibnamefont {Clarke}}, \bibinfo {author} {\bibfnamefont {I.}~\bibnamefont {Siddiqi}}, \bibinfo {author} {\bibfnamefont {A.}~\bibnamefont {Agrawal}},
  \bibinfo {author} {\bibfnamefont {A.~V.}\ \bibnamefont {Dixit}}, \bibinfo {author} {\bibfnamefont {M.~H.}\ \bibnamefont {Awida}}, \bibinfo {author} {\bibfnamefont {A.~S.}\ \bibnamefont {Chou}}, \bibinfo {author} {\bibfnamefont {M.}~\bibnamefont {Hollister}}, \bibinfo {author} {\bibfnamefont {S.}~\bibnamefont {Knirck}}, \bibinfo {author} {\bibfnamefont {A.}~\bibnamefont {Sonnenschein}}, \bibinfo {author} {\bibfnamefont {W.}~\bibnamefont {Wester}}, \bibinfo {author} {\bibfnamefont {J.~R.}\ \bibnamefont {Gleason}}, \bibinfo {author} {\bibfnamefont {A.~T.}\ \bibnamefont {Hipp}}, \bibinfo {author} {\bibfnamefont {S.}~\bibnamefont {Jois}}, \bibinfo {author} {\bibfnamefont {P.}~\bibnamefont {Sikivie}}, \bibinfo {author} {\bibfnamefont {N.~S.}\ \bibnamefont {Sullivan}}, \bibinfo {author} {\bibfnamefont {D.~B.}\ \bibnamefont {Tanner}}, \bibinfo {author} {\bibfnamefont {E.}~\bibnamefont {Lentz}}, \bibinfo {author} {\bibfnamefont {R.}~\bibnamefont {Khatiwada}}, \bibinfo {author} {\bibfnamefont {G.}~\bibnamefont
  {Carosi}}, \bibinfo {author} {\bibfnamefont {N.}~\bibnamefont {Robertson}}, \bibinfo {author} {\bibfnamefont {N.}~\bibnamefont {Woollett}}, \bibinfo {author} {\bibfnamefont {L.~D.}\ \bibnamefont {Duffy}}, \bibinfo {author} {\bibfnamefont {C.}~\bibnamefont {Boutan}}, \bibinfo {author} {\bibfnamefont {M.}~\bibnamefont {Jones}}, \bibinfo {author} {\bibfnamefont {B.~H.}\ \bibnamefont {LaRoque}}, \bibinfo {author} {\bibfnamefont {N.~S.}\ \bibnamefont {Oblath}}, \bibinfo {author} {\bibfnamefont {M.~S.}\ \bibnamefont {Taubman}}, \bibinfo {author} {\bibfnamefont {E.~J.}\ \bibnamefont {Daw}}, \bibinfo {author} {\bibfnamefont {M.~G.}\ \bibnamefont {Perry}}, \bibinfo {author} {\bibfnamefont {J.~H.}\ \bibnamefont {Buckley}}, \bibinfo {author} {\bibfnamefont {C.}~\bibnamefont {Gaikwad}}, \bibinfo {author} {\bibfnamefont {J.}~\bibnamefont {Hoffman}}, \bibinfo {author} {\bibfnamefont {K.~W.}\ \bibnamefont {Murch}}, \bibinfo {author} {\bibfnamefont {M.}~\bibnamefont {Goryachev}}, \bibinfo {author} {\bibfnamefont {B.~T.}\
  \bibnamefont {McAllister}}, \bibinfo {author} {\bibfnamefont {A.}~\bibnamefont {Quiskamp}}, \bibinfo {author} {\bibfnamefont {C.}~\bibnamefont {Thomson}}, \ and\ \bibinfo {author} {\bibfnamefont {M.~E.}\ \bibnamefont {Tobar}} (\bibinfo {collaboration} {ADMX Collaboration}),\ }\href {\doibase 10.1103/PhysRevLett.127.261803} {\bibfield  {journal} {\bibinfo  {journal} {Phys. Rev. Lett.}\ }\textbf {\bibinfo {volume} {127}},\ \bibinfo {pages} {261803} (\bibinfo {year} {2021}{\natexlab{a}})}\BibitemShut {NoStop}%
\bibitem [{\citenamefont {Duffy}\ and\ \citenamefont {Sikivie}(2008)}]{PhysRevD.78.063508}%
  \BibitemOpen
  \bibfield  {author} {\bibinfo {author} {\bibfnamefont {L.~D.}\ \bibnamefont {Duffy}}\ and\ \bibinfo {author} {\bibfnamefont {P.}~\bibnamefont {Sikivie}},\ }\href {\doibase 10.1103/PhysRevD.78.063508} {\bibfield  {journal} {\bibinfo  {journal} {Phys. Rev. D}\ }\textbf {\bibinfo {volume} {78}},\ \bibinfo {pages} {063508} (\bibinfo {year} {2008})}\BibitemShut {NoStop}%
\bibitem [{\citenamefont {Duffy}(2006)}]{duffy_2006}%
  \BibitemOpen
  \bibfield  {author} {\bibinfo {author} {\bibfnamefont {L.~D.}\ \bibnamefont {Duffy}},\ }\emph {\bibinfo {title} {High resolution search for dark matter axions in milky way halo substructure}},\ \href@noop {} {Ph.D. thesis},\ \bibinfo  {school} {University of Florida} (\bibinfo {year} {2006})\BibitemShut {NoStop}%
\bibitem [{\citenamefont {Duffy}\ \emph {et~al.}(2005)\citenamefont {Duffy}, \citenamefont {Sikivie}, \citenamefont {Tanner}, \citenamefont {Asztalos}, \citenamefont {Hagmann}, \citenamefont {Kinion}, \citenamefont {Rosenberg}, \citenamefont {van Bibber}, \citenamefont {Yu},\ and\ \citenamefont {Bradley}}]{Duffy_2005}%
  \BibitemOpen
  \bibfield  {author} {\bibinfo {author} {\bibfnamefont {L.}~\bibnamefont {Duffy}}, \bibinfo {author} {\bibfnamefont {P.}~\bibnamefont {Sikivie}}, \bibinfo {author} {\bibfnamefont {D.~B.}\ \bibnamefont {Tanner}}, \bibinfo {author} {\bibfnamefont {S.}~\bibnamefont {Asztalos}}, \bibinfo {author} {\bibfnamefont {C.}~\bibnamefont {Hagmann}}, \bibinfo {author} {\bibfnamefont {D.}~\bibnamefont {Kinion}}, \bibinfo {author} {\bibfnamefont {L.~J.}\ \bibnamefont {Rosenberg}}, \bibinfo {author} {\bibfnamefont {K.}~\bibnamefont {van Bibber}}, \bibinfo {author} {\bibfnamefont {D.}~\bibnamefont {Yu}}, \ and\ \bibinfo {author} {\bibfnamefont {R.~F.}\ \bibnamefont {Bradley}},\ }\href {\doibase 10.1103/physrevlett.95.091304} {\bibfield  {journal} {\bibinfo  {journal} {Physical Review Letters}\ }\textbf {\bibinfo {volume} {95}} (\bibinfo {year} {2005}),\ 10.1103/physrevlett.95.091304}\BibitemShut {NoStop}%
\bibitem [{\citenamefont {Stiff}\ and\ \citenamefont {Widrow}(2003)}]{Stiff2003}%
  \BibitemOpen
  \bibfield  {author} {\bibinfo {author} {\bibfnamefont {D.}~\bibnamefont {Stiff}}\ and\ \bibinfo {author} {\bibfnamefont {L.~M.}\ \bibnamefont {Widrow}},\ }\href {\doibase 10.1103/PhysRevLett.90.211301} {\bibfield  {journal} {\bibinfo  {journal} {Phys. Rev. Lett.}\ }\textbf {\bibinfo {volume} {90}},\ \bibinfo {pages} {211301} (\bibinfo {year} {2003})}\BibitemShut {NoStop}%
\bibitem [{\citenamefont {Hoskins}\ \emph {et~al.}(2016)\citenamefont {Hoskins}, \citenamefont {Crisosto}, \citenamefont {Gleason}, \citenamefont {Sikivie}, \citenamefont {Stern}, \citenamefont {Sullivan}, \citenamefont {Tanner}, \citenamefont {Boutan}, \citenamefont {Hotz}, \citenamefont {Khatiwada}, \citenamefont {Lyapustin}, \citenamefont {Malagon}, \citenamefont {Ottens}, \citenamefont {Rosenberg}, \citenamefont {Rybka}, \citenamefont {Sloan}, \citenamefont {Wagner}, \citenamefont {Will}, \citenamefont {Carosi}, \citenamefont {Carter}, \citenamefont {Duffy}, \citenamefont {Bradley}, \citenamefont {Clarke}, \citenamefont {O'Kelley}, \citenamefont {van Bibber},\ and\ \citenamefont {Daw}}]{Hoskins_2016}%
  \BibitemOpen
  \bibfield  {author} {\bibinfo {author} {\bibfnamefont {J.}~\bibnamefont {Hoskins}}, \bibinfo {author} {\bibfnamefont {N.}~\bibnamefont {Crisosto}}, \bibinfo {author} {\bibfnamefont {J.}~\bibnamefont {Gleason}}, \bibinfo {author} {\bibfnamefont {P.}~\bibnamefont {Sikivie}}, \bibinfo {author} {\bibfnamefont {I.}~\bibnamefont {Stern}}, \bibinfo {author} {\bibfnamefont {N.}~\bibnamefont {Sullivan}}, \bibinfo {author} {\bibfnamefont {D.}~\bibnamefont {Tanner}}, \bibinfo {author} {\bibfnamefont {C.}~\bibnamefont {Boutan}}, \bibinfo {author} {\bibfnamefont {M.}~\bibnamefont {Hotz}}, \bibinfo {author} {\bibfnamefont {R.}~\bibnamefont {Khatiwada}}, \bibinfo {author} {\bibfnamefont {D.}~\bibnamefont {Lyapustin}}, \bibinfo {author} {\bibfnamefont {A.}~\bibnamefont {Malagon}}, \bibinfo {author} {\bibfnamefont {R.}~\bibnamefont {Ottens}}, \bibinfo {author} {\bibfnamefont {L.}~\bibnamefont {Rosenberg}}, \bibinfo {author} {\bibfnamefont {G.}~\bibnamefont {Rybka}}, \bibinfo {author} {\bibfnamefont {J.}~\bibnamefont
  {Sloan}}, \bibinfo {author} {\bibfnamefont {A.}~\bibnamefont {Wagner}}, \bibinfo {author} {\bibfnamefont {D.}~\bibnamefont {Will}}, \bibinfo {author} {\bibfnamefont {G.}~\bibnamefont {Carosi}}, \bibinfo {author} {\bibfnamefont {D.}~\bibnamefont {Carter}}, \bibinfo {author} {\bibfnamefont {L.}~\bibnamefont {Duffy}}, \bibinfo {author} {\bibfnamefont {R.}~\bibnamefont {Bradley}}, \bibinfo {author} {\bibfnamefont {J.}~\bibnamefont {Clarke}}, \bibinfo {author} {\bibfnamefont {S.}~\bibnamefont {O'Kelley}}, \bibinfo {author} {\bibfnamefont {K.}~\bibnamefont {van Bibber}}, \ and\ \bibinfo {author} {\bibfnamefont {E.}~\bibnamefont {Daw}},\ }\href {\doibase 10.1103/physrevd.94.082001} {\bibfield  {journal} {\bibinfo  {journal} {Physical Review D}\ }\textbf {\bibinfo {volume} {94}} (\bibinfo {year} {2016}),\ 10.1103/physrevd.94.082001}\BibitemShut {NoStop}%
\bibitem [{\citenamefont {Sikivie}(1998)}]{Sikivie_1998}%
  \BibitemOpen
  \bibfield  {author} {\bibinfo {author} {\bibfnamefont {P.}~\bibnamefont {Sikivie}},\ }\href {\doibase https://doi.org/10.1016/S0370-2693(98)00595-4} {\bibfield  {journal} {\bibinfo  {journal} {Physics Letters B}\ }\textbf {\bibinfo {volume} {432}},\ \bibinfo {pages} {139} (\bibinfo {year} {1998})}\BibitemShut {NoStop}%
\bibitem [{\citenamefont {Sikivie}(1999)}]{Sikivie1999}%
  \BibitemOpen
  \bibfield  {author} {\bibinfo {author} {\bibfnamefont {P.}~\bibnamefont {Sikivie}},\ }\href {\doibase 10.1103/PhysRevD.60.063501} {\bibfield  {journal} {\bibinfo  {journal} {Phys. Rev. D}\ }\textbf {\bibinfo {volume} {60}},\ \bibinfo {pages} {063501} (\bibinfo {year} {1999})}\BibitemShut {NoStop}%
\bibitem [{\citenamefont {Chakrabarty}\ \emph {et~al.}(2021)\citenamefont {Chakrabarty}, \citenamefont {Han}, \citenamefont {Gonzalez},\ and\ \citenamefont {Sikivie}}]{CHAKRABARTY2021100838}%
  \BibitemOpen
  \bibfield  {author} {\bibinfo {author} {\bibfnamefont {S.~S.}\ \bibnamefont {Chakrabarty}}, \bibinfo {author} {\bibfnamefont {Y.}~\bibnamefont {Han}}, \bibinfo {author} {\bibfnamefont {A.~H.}\ \bibnamefont {Gonzalez}}, \ and\ \bibinfo {author} {\bibfnamefont {P.}~\bibnamefont {Sikivie}},\ }\href {\doibase https://doi.org/10.1016/j.dark.2021.100838} {\bibfield  {journal} {\bibinfo  {journal} {Physics of the Dark Universe}\ }\textbf {\bibinfo {volume} {33}},\ \bibinfo {pages} {100838} (\bibinfo {year} {2021})}\BibitemShut {NoStop}%
\bibitem [{\citenamefont {Sikivie}(2003)}]{sikivie2003evidence}%
  \BibitemOpen
  \bibfield  {author} {\bibinfo {author} {\bibfnamefont {P.}~\bibnamefont {Sikivie}},\ }\href@noop {} {\bibfield  {journal} {\bibinfo  {journal} {Physics Letters B}\ }\textbf {\bibinfo {volume} {567}},\ \bibinfo {pages} {1} (\bibinfo {year} {2003})}\BibitemShut {NoStop}%
\bibitem [{\citenamefont {Freese}\ \emph {et~al.}(2005)\citenamefont {Freese}, \citenamefont {Gondolo},\ and\ \citenamefont {Newberg}}]{PhysRevD.71.043516}%
  \BibitemOpen
  \bibfield  {author} {\bibinfo {author} {\bibfnamefont {K.}~\bibnamefont {Freese}}, \bibinfo {author} {\bibfnamefont {P.}~\bibnamefont {Gondolo}}, \ and\ \bibinfo {author} {\bibfnamefont {H.~J.}\ \bibnamefont {Newberg}},\ }\href {\doibase 10.1103/PhysRevD.71.043516} {\bibfield  {journal} {\bibinfo  {journal} {Phys. Rev. D}\ }\textbf {\bibinfo {volume} {71}},\ \bibinfo {pages} {043516} (\bibinfo {year} {2005})}\BibitemShut {NoStop}%
\bibitem [{\citenamefont {Freese}\ \emph {et~al.}(2004)\citenamefont {Freese}, \citenamefont {Gondolo}, \citenamefont {Newberg},\ and\ \citenamefont {Lewis}}]{PhysRevLett.92.111301}%
  \BibitemOpen
  \bibfield  {author} {\bibinfo {author} {\bibfnamefont {K.}~\bibnamefont {Freese}}, \bibinfo {author} {\bibfnamefont {P.}~\bibnamefont {Gondolo}}, \bibinfo {author} {\bibfnamefont {H.~J.}\ \bibnamefont {Newberg}}, \ and\ \bibinfo {author} {\bibfnamefont {M.}~\bibnamefont {Lewis}},\ }\href {\doibase 10.1103/PhysRevLett.92.111301} {\bibfield  {journal} {\bibinfo  {journal} {Phys. Rev. Lett.}\ }\textbf {\bibinfo {volume} {92}},\ \bibinfo {pages} {111301} (\bibinfo {year} {2004})}\BibitemShut {NoStop}%
\bibitem [{\citenamefont {Sadashivajois}(2020)}]{Ram_Thesis}%
  \BibitemOpen
  \bibfield  {author} {\bibinfo {author} {\bibfnamefont {S.}~\bibnamefont {Sadashivajois}},\ }\emph {\bibinfo {title} {A Search for Relic Axions and Their Frequency Modulation}},\ \href@noop {} {Ph.D. thesis},\ \bibinfo  {school} {University of Florida} (\bibinfo {year} {2020})\BibitemShut {NoStop}%
\bibitem [{\citenamefont {Read}(2014)}]{Read2014}%
  \BibitemOpen
  \bibfield  {author} {\bibinfo {author} {\bibfnamefont {J.~I.}\ \bibnamefont {Read}},\ }\href {\doibase 10.1088/0954-3899/41/6/063101} {\bibfield  {journal} {\bibinfo  {journal} {Journal of Physics G: Nuclear and Particle Physics}\ }\textbf {\bibinfo {volume} {41}},\ \bibinfo {pages} {063101} (\bibinfo {year} {2014})}\BibitemShut {NoStop}%
\bibitem [{\citenamefont {Khatiwada}\ \emph {et~al.}(2020)\citenamefont {Khatiwada}, \citenamefont {Bowring}, \citenamefont {Chou}, \citenamefont {Sonnenschein}, \citenamefont {Wester}, \citenamefont {Mitchell}, \citenamefont {Braine}, \citenamefont {Bartram}, \citenamefont {Cervantes}, \citenamefont {Crisosto}, \citenamefont {Du}, \citenamefont {Rosenberg}, \citenamefont {Rybka}, \citenamefont {Yang}, \citenamefont {Will}, \citenamefont {Kimes}, \citenamefont {Carosi}, \citenamefont {Woollett}, \citenamefont {Durham}, \citenamefont {Duffy}, \citenamefont {Bradley}, \citenamefont {Boutan}, \citenamefont {Jones}, \citenamefont {LaRoque}, \citenamefont {Oblath}, \citenamefont {Taubman}, \citenamefont {Tedeschi}, \citenamefont {Clarke}, \citenamefont {Dove}, \citenamefont {Hashim}, \citenamefont {Siddiqi}, \citenamefont {Stevenson}, \citenamefont {Eddins}, \citenamefont {O'kelley}, \citenamefont {Nawaz}, \citenamefont {Agrawal}, \citenamefont {Dixit}, \citenamefont {Gleason}, \citenamefont {Jois}, \citenamefont
  {Sikivie}, \citenamefont {Sullivan}, \citenamefont {Tanner}, \citenamefont {Solomon}, \citenamefont {Lentz}, \citenamefont {Daw}, \citenamefont {Perry}, \citenamefont {Buckley}, \citenamefont {Harrington}, \citenamefont {Henriksen}, \citenamefont {Murch},\ and\ \citenamefont {Hilton}}]{Khatiwada2020AxionDM}%
  \BibitemOpen
  \bibfield  {author} {\bibinfo {author} {\bibfnamefont {R.}~\bibnamefont {Khatiwada}}, \bibinfo {author} {\bibfnamefont {D.}~\bibnamefont {Bowring}}, \bibinfo {author} {\bibfnamefont {A.}~\bibnamefont {Chou}}, \bibinfo {author} {\bibfnamefont {A.~H.}\ \bibnamefont {Sonnenschein}}, \bibinfo {author} {\bibfnamefont {W.~C.}\ \bibnamefont {Wester}}, \bibinfo {author} {\bibfnamefont {D.~V.}\ \bibnamefont {Mitchell}}, \bibinfo {author} {\bibfnamefont {T.}~\bibnamefont {Braine}}, \bibinfo {author} {\bibfnamefont {C.}~\bibnamefont {Bartram}}, \bibinfo {author} {\bibfnamefont {R.}~\bibnamefont {Cervantes}}, \bibinfo {author} {\bibfnamefont {N.}~\bibnamefont {Crisosto}}, \bibinfo {author} {\bibfnamefont {N.}~\bibnamefont {Du}}, \bibinfo {author} {\bibfnamefont {L.~J.}\ \bibnamefont {Rosenberg}}, \bibinfo {author} {\bibfnamefont {G.}~\bibnamefont {Rybka}}, \bibinfo {author} {\bibfnamefont {J.}~\bibnamefont {Yang}}, \bibinfo {author} {\bibfnamefont {D.~I.}\ \bibnamefont {Will}}, \bibinfo {author} {\bibfnamefont
  {S.}~\bibnamefont {Kimes}}, \bibinfo {author} {\bibfnamefont {G.}~\bibnamefont {Carosi}}, \bibinfo {author} {\bibfnamefont {N.}~\bibnamefont {Woollett}}, \bibinfo {author} {\bibfnamefont {S.}~\bibnamefont {Durham}}, \bibinfo {author} {\bibfnamefont {L.~D.}\ \bibnamefont {Duffy}}, \bibinfo {author} {\bibfnamefont {R.~F.}\ \bibnamefont {Bradley}}, \bibinfo {author} {\bibfnamefont {C.}~\bibnamefont {Boutan}}, \bibinfo {author} {\bibfnamefont {M.}~\bibnamefont {Jones}}, \bibinfo {author} {\bibfnamefont {B.}~\bibnamefont {LaRoque}}, \bibinfo {author} {\bibfnamefont {N.~S.}\ \bibnamefont {Oblath}}, \bibinfo {author} {\bibfnamefont {M.~S.}\ \bibnamefont {Taubman}}, \bibinfo {author} {\bibfnamefont {J.~R.}\ \bibnamefont {Tedeschi}}, \bibinfo {author} {\bibfnamefont {J.~D.}\ \bibnamefont {Clarke}}, \bibinfo {author} {\bibfnamefont {A.}~\bibnamefont {Dove}}, \bibinfo {author} {\bibfnamefont {A.}~\bibnamefont {Hashim}}, \bibinfo {author} {\bibfnamefont {I.}~\bibnamefont {Siddiqi}}, \bibinfo {author} {\bibfnamefont
  {N.}~\bibnamefont {Stevenson}}, \bibinfo {author} {\bibfnamefont {A.}~\bibnamefont {Eddins}}, \bibinfo {author} {\bibfnamefont {S.}~\bibnamefont {O'kelley}}, \bibinfo {author} {\bibfnamefont {S.}~\bibnamefont {Nawaz}}, \bibinfo {author} {\bibfnamefont {A.}~\bibnamefont {Agrawal}}, \bibinfo {author} {\bibfnamefont {A.~V.}\ \bibnamefont {Dixit}}, \bibinfo {author} {\bibfnamefont {J.~R.}\ \bibnamefont {Gleason}}, \bibinfo {author} {\bibfnamefont {S.}~\bibnamefont {Jois}}, \bibinfo {author} {\bibfnamefont {P.}~\bibnamefont {Sikivie}}, \bibinfo {author} {\bibfnamefont {N.~S.}\ \bibnamefont {Sullivan}}, \bibinfo {author} {\bibfnamefont {D.~B.}\ \bibnamefont {Tanner}}, \bibinfo {author} {\bibfnamefont {J.}~\bibnamefont {Solomon}}, \bibinfo {author} {\bibfnamefont {E.~W.}\ \bibnamefont {Lentz}}, \bibinfo {author} {\bibfnamefont {E.~J.}\ \bibnamefont {Daw}}, \bibinfo {author} {\bibfnamefont {M.~G.}\ \bibnamefont {Perry}}, \bibinfo {author} {\bibfnamefont {J.~H.}\ \bibnamefont {Buckley}}, \bibinfo {author}
  {\bibfnamefont {P.~M.}\ \bibnamefont {Harrington}}, \bibinfo {author} {\bibfnamefont {E.~A.}\ \bibnamefont {Henriksen}}, \bibinfo {author} {\bibfnamefont {K.~W.}\ \bibnamefont {Murch}}, \ and\ \bibinfo {author} {\bibfnamefont {G.~C.}\ \bibnamefont {Hilton}},\ }\href {https://api.semanticscholar.org/CorpusID:222090014} {\bibfield  {journal} {\bibinfo  {journal} {The Review of scientific instruments}\ }\textbf {\bibinfo {volume} {92 12}},\ \bibinfo {pages} {124502} (\bibinfo {year} {2020})}\BibitemShut {NoStop}%
\bibitem [{\citenamefont {Bartram}\ \emph {et~al.}(2021{\natexlab{b}})\citenamefont {Bartram}, \citenamefont {Braine}, \citenamefont {Cervantes}, \citenamefont {Crisosto}, \citenamefont {Du}, \citenamefont {Leum}, \citenamefont {Rosenberg}, \citenamefont {Rybka}, \citenamefont {Yang}, \citenamefont {Bowring}, \citenamefont {Chou}, \citenamefont {Khatiwada}, \citenamefont {Sonnenschein}, \citenamefont {Wester}, \citenamefont {Carosi}, \citenamefont {Woollett}, \citenamefont {Duffy}, \citenamefont {Goryachev}, \citenamefont {McAllister}, \citenamefont {Tobar}, \citenamefont {Boutan}, \citenamefont {Jones}, \citenamefont {LaRoque}, \citenamefont {Oblath}, \citenamefont {Taubman}, \citenamefont {Clarke}, \citenamefont {Dove}, \citenamefont {Eddins}, \citenamefont {O'Kelley}, \citenamefont {Nawaz}, \citenamefont {Siddiqi}, \citenamefont {Stevenson}, \citenamefont {Agrawal}, \citenamefont {Dixit}, \citenamefont {Gleason}, \citenamefont {Jois}, \citenamefont {Sikivie}, \citenamefont {Solomon}, \citenamefont
  {Sullivan}, \citenamefont {Tanner}, \citenamefont {Lentz}, \citenamefont {Daw}, \citenamefont {Perry}, \citenamefont {Buckley}, \citenamefont {Harrington}, \citenamefont {Henriksen},\ and\ \citenamefont {Murch}}]{bartram2021axion}%
  \BibitemOpen
  \bibfield  {author} {\bibinfo {author} {\bibfnamefont {C.}~\bibnamefont {Bartram}}, \bibinfo {author} {\bibfnamefont {T.}~\bibnamefont {Braine}}, \bibinfo {author} {\bibfnamefont {R.}~\bibnamefont {Cervantes}}, \bibinfo {author} {\bibfnamefont {N.}~\bibnamefont {Crisosto}}, \bibinfo {author} {\bibfnamefont {N.}~\bibnamefont {Du}}, \bibinfo {author} {\bibfnamefont {G.}~\bibnamefont {Leum}}, \bibinfo {author} {\bibfnamefont {L.~J.}\ \bibnamefont {Rosenberg}}, \bibinfo {author} {\bibfnamefont {G.}~\bibnamefont {Rybka}}, \bibinfo {author} {\bibfnamefont {J.}~\bibnamefont {Yang}}, \bibinfo {author} {\bibfnamefont {D.}~\bibnamefont {Bowring}}, \bibinfo {author} {\bibfnamefont {A.~S.}\ \bibnamefont {Chou}}, \bibinfo {author} {\bibfnamefont {R.}~\bibnamefont {Khatiwada}}, \bibinfo {author} {\bibfnamefont {A.}~\bibnamefont {Sonnenschein}}, \bibinfo {author} {\bibfnamefont {W.}~\bibnamefont {Wester}}, \bibinfo {author} {\bibfnamefont {G.}~\bibnamefont {Carosi}}, \bibinfo {author} {\bibfnamefont {N.}~\bibnamefont
  {Woollett}}, \bibinfo {author} {\bibfnamefont {L.~D.}\ \bibnamefont {Duffy}}, \bibinfo {author} {\bibfnamefont {M.}~\bibnamefont {Goryachev}}, \bibinfo {author} {\bibfnamefont {B.}~\bibnamefont {McAllister}}, \bibinfo {author} {\bibfnamefont {M.~E.}\ \bibnamefont {Tobar}}, \bibinfo {author} {\bibfnamefont {C.}~\bibnamefont {Boutan}}, \bibinfo {author} {\bibfnamefont {M.}~\bibnamefont {Jones}}, \bibinfo {author} {\bibfnamefont {B.~H.}\ \bibnamefont {LaRoque}}, \bibinfo {author} {\bibfnamefont {N.~S.}\ \bibnamefont {Oblath}}, \bibinfo {author} {\bibfnamefont {M.~S.}\ \bibnamefont {Taubman}}, \bibinfo {author} {\bibfnamefont {J.}~\bibnamefont {Clarke}}, \bibinfo {author} {\bibfnamefont {A.}~\bibnamefont {Dove}}, \bibinfo {author} {\bibfnamefont {A.}~\bibnamefont {Eddins}}, \bibinfo {author} {\bibfnamefont {S.~R.}\ \bibnamefont {O'Kelley}}, \bibinfo {author} {\bibfnamefont {S.}~\bibnamefont {Nawaz}}, \bibinfo {author} {\bibfnamefont {I.}~\bibnamefont {Siddiqi}}, \bibinfo {author} {\bibfnamefont
  {N.}~\bibnamefont {Stevenson}}, \bibinfo {author} {\bibfnamefont {A.}~\bibnamefont {Agrawal}}, \bibinfo {author} {\bibfnamefont {A.~V.}\ \bibnamefont {Dixit}}, \bibinfo {author} {\bibfnamefont {J.~R.}\ \bibnamefont {Gleason}}, \bibinfo {author} {\bibfnamefont {S.}~\bibnamefont {Jois}}, \bibinfo {author} {\bibfnamefont {P.}~\bibnamefont {Sikivie}}, \bibinfo {author} {\bibfnamefont {J.~A.}\ \bibnamefont {Solomon}}, \bibinfo {author} {\bibfnamefont {N.~S.}\ \bibnamefont {Sullivan}}, \bibinfo {author} {\bibfnamefont {D.~B.}\ \bibnamefont {Tanner}}, \bibinfo {author} {\bibfnamefont {E.}~\bibnamefont {Lentz}}, \bibinfo {author} {\bibfnamefont {E.~J.}\ \bibnamefont {Daw}}, \bibinfo {author} {\bibfnamefont {M.~G.}\ \bibnamefont {Perry}}, \bibinfo {author} {\bibfnamefont {J.~H.}\ \bibnamefont {Buckley}}, \bibinfo {author} {\bibfnamefont {P.~M.}\ \bibnamefont {Harrington}}, \bibinfo {author} {\bibfnamefont {E.~A.}\ \bibnamefont {Henriksen}}, \ and\ \bibinfo {author} {\bibfnamefont {K.~W.}\ \bibnamefont {Murch}}
  (\bibinfo {collaboration} {ADMX Collaboration}),\ }\href {\doibase 10.1103/PhysRevD.103.032002} {\bibfield  {journal} {\bibinfo  {journal} {Phys. Rev. D}\ }\textbf {\bibinfo {volume} {103}},\ \bibinfo {pages} {032002} (\bibinfo {year} {2021}{\natexlab{b}})}\BibitemShut {NoStop}%
\bibitem [{\citenamefont {{Hatridge}}\ \emph {et~al.}(2011)\citenamefont {{Hatridge}}, \citenamefont {{Vijay}}, \citenamefont {{Slichter}}, \citenamefont {{Clarke}},\ and\ \citenamefont {{Siddiqi}}}]{JPA1}%
  \BibitemOpen
  \bibfield  {author} {\bibinfo {author} {\bibfnamefont {M.}~\bibnamefont {{Hatridge}}}, \bibinfo {author} {\bibfnamefont {R.}~\bibnamefont {{Vijay}}}, \bibinfo {author} {\bibfnamefont {D.~H.}\ \bibnamefont {{Slichter}}}, \bibinfo {author} {\bibfnamefont {J.}~\bibnamefont {{Clarke}}}, \ and\ \bibinfo {author} {\bibfnamefont {I.}~\bibnamefont {{Siddiqi}}},\ }\href {\doibase 10.1103/PhysRevB.83.134501} {\bibfield  {journal} {\bibinfo  {journal} {\prb}\ }\textbf {\bibinfo {volume} {83}},\ \bibinfo {eid} {134501} (\bibinfo {year} {2011})},\ \Eprint {http://arxiv.org/abs/1003.2466} {arXiv:1003.2466 [cond-mat.mes-hall]} \BibitemShut {NoStop}%
\bibitem [{\citenamefont {Boutin}\ \emph {et~al.}(2017)\citenamefont {Boutin}, \citenamefont {Toyli}, \citenamefont {Venkatramani}, \citenamefont {Eddins}, \citenamefont {Siddiqi},\ and\ \citenamefont {Blais}}]{JPA2}%
  \BibitemOpen
  \bibfield  {author} {\bibinfo {author} {\bibfnamefont {S.}~\bibnamefont {Boutin}}, \bibinfo {author} {\bibfnamefont {D.~M.}\ \bibnamefont {Toyli}}, \bibinfo {author} {\bibfnamefont {A.~V.}\ \bibnamefont {Venkatramani}}, \bibinfo {author} {\bibfnamefont {A.~W.}\ \bibnamefont {Eddins}}, \bibinfo {author} {\bibfnamefont {I.}~\bibnamefont {Siddiqi}}, \ and\ \bibinfo {author} {\bibfnamefont {A.}~\bibnamefont {Blais}},\ }\href {\doibase 10.1103/PhysRevApplied.8.054030} {\bibfield  {journal} {\bibinfo  {journal} {Phys. Rev. Applied}\ }\textbf {\bibinfo {volume} {8}},\ \bibinfo {pages} {054030} (\bibinfo {year} {2017})}\BibitemShut {NoStop}%
\bibitem [{LNF(2022)}]{LNF}%
  \BibitemOpen
  \href {https://lownoisefactory.com/} {} (\bibinfo {year} {2022})\BibitemShut {NoStop}%
\bibitem [{CST(2023)}]{CST}%
  \BibitemOpen
  \href@noop {} {\enquote {\bibinfo {title} {{CST Studio Suite}},}\ }\bibinfo {howpublished} {\url{https://www.3ds.com/products-services/simulia/products/cst-studio-suite/}} (\bibinfo {year} {2023}),\ \bibinfo {note} {accessed: 2023-02-27}\BibitemShut {NoStop}%
\bibitem [{\citenamefont {Duffy}\ \emph {et~al.}(2006)\citenamefont {Duffy}, \citenamefont {Sikivie}, \citenamefont {Tanner}, \citenamefont {Asztalos}, \citenamefont {Hagmann}, \citenamefont {Kinion}, \citenamefont {Rosenberg}, \citenamefont {van Bibber}, \citenamefont {Yu},\ and\ \citenamefont {Bradley}}]{Leanne_2006}%
  \BibitemOpen
  \bibfield  {author} {\bibinfo {author} {\bibfnamefont {L.~D.}\ \bibnamefont {Duffy}}, \bibinfo {author} {\bibfnamefont {P.}~\bibnamefont {Sikivie}}, \bibinfo {author} {\bibfnamefont {D.~B.}\ \bibnamefont {Tanner}}, \bibinfo {author} {\bibfnamefont {S.~J.}\ \bibnamefont {Asztalos}}, \bibinfo {author} {\bibfnamefont {C.}~\bibnamefont {Hagmann}}, \bibinfo {author} {\bibfnamefont {D.}~\bibnamefont {Kinion}}, \bibinfo {author} {\bibfnamefont {L.~J.}\ \bibnamefont {Rosenberg}}, \bibinfo {author} {\bibfnamefont {K.}~\bibnamefont {van Bibber}}, \bibinfo {author} {\bibfnamefont {D.~B.}\ \bibnamefont {Yu}}, \ and\ \bibinfo {author} {\bibfnamefont {R.~F.}\ \bibnamefont {Bradley}},\ }\href {\doibase 10.1103/PhysRevD.74.012006} {\bibfield  {journal} {\bibinfo  {journal} {Phys. Rev. D}\ }\textbf {\bibinfo {volume} {74}},\ \bibinfo {pages} {012006} (\bibinfo {year} {2006})}\BibitemShut {NoStop}%
\bibitem [{\citenamefont {Brubaker}\ \emph {et~al.}(2017)\citenamefont {Brubaker}, \citenamefont {Zhong}, \citenamefont {Lamoreaux}, \citenamefont {Lehnert},\ and\ \citenamefont {van Bibber}}]{haystac_analysis}%
  \BibitemOpen
  \bibfield  {author} {\bibinfo {author} {\bibfnamefont {B.~M.}\ \bibnamefont {Brubaker}}, \bibinfo {author} {\bibfnamefont {L.}~\bibnamefont {Zhong}}, \bibinfo {author} {\bibfnamefont {S.~K.}\ \bibnamefont {Lamoreaux}}, \bibinfo {author} {\bibfnamefont {K.~W.}\ \bibnamefont {Lehnert}}, \ and\ \bibinfo {author} {\bibfnamefont {K.~A.}\ \bibnamefont {van Bibber}},\ }\href {\doibase 10.1103/PhysRevD.96.123008} {\bibfield  {journal} {\bibinfo  {journal} {Phys. Rev. D}\ }\textbf {\bibinfo {volume} {96}},\ \bibinfo {pages} {123008} (\bibinfo {year} {2017})}\BibitemShut {NoStop}%
\bibitem [{\citenamefont {Daw}(1998)}]{Daw_thesis}%
  \BibitemOpen
  \bibfield  {author} {\bibinfo {author} {\bibfnamefont {E.~J.}\ \bibnamefont {Daw}},\ }\emph {\bibinfo {title} {A Search for Halo Axions}},\ \href@noop {} {Ph.D. thesis},\ \bibinfo  {school} {Massachusetts Institute of Technology}, \bibinfo {address} {Cambridge, MA} (\bibinfo {year} {1998})\BibitemShut {NoStop}%
\bibitem [{\citenamefont {Asztalos}\ \emph {et~al.}(2010)\citenamefont {Asztalos}, \citenamefont {Carosi}, \citenamefont {Hagmann}, \citenamefont {Kinion}, \citenamefont {van Bibber}, \citenamefont {Hotz}, \citenamefont {Rosenberg}, \citenamefont {Rybka}, \citenamefont {Hoskins}, \citenamefont {Hwang}, \citenamefont {Sikivie}, \citenamefont {Tanner}, \citenamefont {Bradley},\ and\ \citenamefont {Clarke}}]{ADMX_2010}%
  \BibitemOpen
  \bibfield  {author} {\bibinfo {author} {\bibfnamefont {S.~J.}\ \bibnamefont {Asztalos}}, \bibinfo {author} {\bibfnamefont {G.}~\bibnamefont {Carosi}}, \bibinfo {author} {\bibfnamefont {C.}~\bibnamefont {Hagmann}}, \bibinfo {author} {\bibfnamefont {D.}~\bibnamefont {Kinion}}, \bibinfo {author} {\bibfnamefont {K.}~\bibnamefont {van Bibber}}, \bibinfo {author} {\bibfnamefont {M.}~\bibnamefont {Hotz}}, \bibinfo {author} {\bibfnamefont {L.~J.}\ \bibnamefont {Rosenberg}}, \bibinfo {author} {\bibfnamefont {G.}~\bibnamefont {Rybka}}, \bibinfo {author} {\bibfnamefont {J.}~\bibnamefont {Hoskins}}, \bibinfo {author} {\bibfnamefont {J.}~\bibnamefont {Hwang}}, \bibinfo {author} {\bibfnamefont {P.}~\bibnamefont {Sikivie}}, \bibinfo {author} {\bibfnamefont {D.~B.}\ \bibnamefont {Tanner}}, \bibinfo {author} {\bibfnamefont {R.}~\bibnamefont {Bradley}}, \ and\ \bibinfo {author} {\bibfnamefont {J.}~\bibnamefont {Clarke}},\ }\href {\doibase 10.1103/PhysRevLett.104.041301} {\bibfield  {journal} {\bibinfo  {journal} {Phys. Rev.
  Lett.}\ }\textbf {\bibinfo {volume} {104}},\ \bibinfo {pages} {041301} (\bibinfo {year} {2010})}\BibitemShut {NoStop}%
\bibitem [{\citenamefont {Asztalos}\ \emph {et~al.}(2001)\citenamefont {Asztalos}, \citenamefont {Daw}, \citenamefont {Peng}, \citenamefont {Rosenberg}, \citenamefont {Hagmann}, \citenamefont {Kinion}, \citenamefont {Stoeffl}, \citenamefont {van Bibber}, \citenamefont {Sikivie}, \citenamefont {Sullivan}, \citenamefont {Tanner}, \citenamefont {Nezrick}, \citenamefont {Turner}, \citenamefont {Moltz}, \citenamefont {Powell}, \citenamefont {Andr\'e}, \citenamefont {Clarke}, \citenamefont {M\"uck},\ and\ \citenamefont {Bradley}}]{MC_limit}%
  \BibitemOpen
  \bibfield  {author} {\bibinfo {author} {\bibfnamefont {S.}~\bibnamefont {Asztalos}}, \bibinfo {author} {\bibfnamefont {E.}~\bibnamefont {Daw}}, \bibinfo {author} {\bibfnamefont {H.}~\bibnamefont {Peng}}, \bibinfo {author} {\bibfnamefont {L.~J.}\ \bibnamefont {Rosenberg}}, \bibinfo {author} {\bibfnamefont {C.}~\bibnamefont {Hagmann}}, \bibinfo {author} {\bibfnamefont {D.}~\bibnamefont {Kinion}}, \bibinfo {author} {\bibfnamefont {W.}~\bibnamefont {Stoeffl}}, \bibinfo {author} {\bibfnamefont {K.}~\bibnamefont {van Bibber}}, \bibinfo {author} {\bibfnamefont {P.}~\bibnamefont {Sikivie}}, \bibinfo {author} {\bibfnamefont {N.~S.}\ \bibnamefont {Sullivan}}, \bibinfo {author} {\bibfnamefont {D.~B.}\ \bibnamefont {Tanner}}, \bibinfo {author} {\bibfnamefont {F.}~\bibnamefont {Nezrick}}, \bibinfo {author} {\bibfnamefont {M.~S.}\ \bibnamefont {Turner}}, \bibinfo {author} {\bibfnamefont {D.~M.}\ \bibnamefont {Moltz}}, \bibinfo {author} {\bibfnamefont {J.}~\bibnamefont {Powell}}, \bibinfo {author} {\bibfnamefont {M.-O.}\
  \bibnamefont {Andr\'e}}, \bibinfo {author} {\bibfnamefont {J.}~\bibnamefont {Clarke}}, \bibinfo {author} {\bibfnamefont {M.}~\bibnamefont {M\"uck}}, \ and\ \bibinfo {author} {\bibfnamefont {R.~F.}\ \bibnamefont {Bradley}},\ }\href {\doibase 10.1103/PhysRevD.64.092003} {\bibfield  {journal} {\bibinfo  {journal} {Phys. Rev. D}\ }\textbf {\bibinfo {volume} {64}},\ \bibinfo {pages} {092003} (\bibinfo {year} {2001})}\BibitemShut {NoStop}%
\bibitem [{\citenamefont {Turner}(1990)}]{Turner_1990}%
  \BibitemOpen
  \bibfield  {author} {\bibinfo {author} {\bibfnamefont {M.~S.}\ \bibnamefont {Turner}},\ }\href {\doibase 10.1103/PhysRevD.42.3572} {\bibfield  {journal} {\bibinfo  {journal} {Phys. Rev. D}\ }\textbf {\bibinfo {volume} {42}},\ \bibinfo {pages} {3572} (\bibinfo {year} {1990})}\BibitemShut {NoStop}%
\bibitem [{\citenamefont {{Ding}}\ \emph {et~al.}(2019)\citenamefont {{Ding}}, \citenamefont {{Zhu}},\ and\ \citenamefont {{Liu}}}]{Ding_2019_LSR}%
  \BibitemOpen
  \bibfield  {author} {\bibinfo {author} {\bibfnamefont {P.-J.}\ \bibnamefont {{Ding}}}, \bibinfo {author} {\bibfnamefont {Z.}~\bibnamefont {{Zhu}}}, \ and\ \bibinfo {author} {\bibfnamefont {J.-C.}\ \bibnamefont {{Liu}}},\ }\href {\doibase 10.1088/1674-4527/19/5/68} {\bibfield  {journal} {\bibinfo  {journal} {Research in Astronomy and Astrophysics}\ }\textbf {\bibinfo {volume} {19}},\ \bibinfo {eid} {068} (\bibinfo {year} {2019})}\BibitemShut {NoStop}%
\end{thebibliography}%
\end{document}